\documentclass[11pt,a4paper]{article}
\usepackage[margin=1in]{geometry}
\usepackage[T1]{fontenc}
\usepackage[utf8]{inputenc}
\usepackage{lmodern}
\usepackage{amsmath,amssymb,amsthm,bm,mathtools}
\usepackage{booktabs,tabularx,array}
\usepackage{enumitem,microtype}
\usepackage{graphicx}
\usepackage[section]{placeins}
\usepackage{xcolor}
\usepackage{hyperref}
\hypersetup{colorlinks=true,linkcolor=blue!45!black,citecolor=blue!45!black,urlcolor=blue!45!black,
 pdftitle={An Exploratory Pilot Analysis of Paired PD-L1 and Siglec-15 SPR Sensorgrams},pdfauthor={Pengfei Wang}}
\numberwithin{equation}{section}
\theoremstyle{plain}
\newtheorem{proposition}{Proposition}[section]
\newtheorem{lemma}[proposition]{Lemma}

\theoremstyle{remark}

\newcommand{\Var}{\operatorname{Var}}

\newcommand{\R}{\mathbb R}
\newcommand{\Med}{\operatorname{Med}}

\newcommand{\AUC}{\operatorname{AUC}}

\newcolumntype{L}{>{\raggedright\arraybackslash}X}
\setlist{itemsep=3pt,topsep=5pt}
\renewcommand{\arraystretch}{1.13}
\title{An Exploratory Pilot Analysis of Paired\\PD-L1 and Siglec-15 SPR Sensorgrams:\\Retained Response, Early Shape, and Complementarity}
\author{Pengfei Wang\\[3pt]
{\small Mathematics and Mathematics Education}\\
{\small National Institute of Education}\\
{\small Nanyang Technological University, Singapore}}
\date{7 November 2026}
\begin{document}
\maketitle
\begin{abstract}
We conduct a retrospective exploratory pilot analysis of paired PD-L1 and Siglec-15 surface plasmon resonance (SPR) trajectories, examining retained response, dual-marker fusion, and additional early-shape information. Numerical matching links the original exports to previously quality-selected data: 69 files represent 68 distinct source column pairs from 10 cancer and 10 healthy study patients. Existing exclusions and truncations are preserved. All 20 patients contribute to development and exploration; no independent validation cohort is available. Because historical per-scan manual alignment offsets are unavailable, the numerical comparisons are conditional on separately documented candidate response-transition intervals rather than verified injection-event times. Patient-level paired comparisons, leave-pair-out internal assessment, empirical false-positive budgets, and timing sensitivity are implemented in reproducible Python code. Under candidate alignment, the dual-marker retained-response sum has an apparent AUC of 0.570, with a fixed-score bootstrap 95\% interval of [0.300, 0.820]. In the nested 13-patient shape subset, retention alone and the fixed 85:15 retention--shape score both have AUC 0.619; the small paired increment changes direction across the $-1,0,+1$ minute shift scenarios. At a same-sample budget of two false positives among ten controls, dual-marker retention identifies four cancer cases, compared with two for PD-L1 and three for Siglec-15. Analytical bounds describe feature stability under stated measurement assumptions but do not identify the cause of weak discrimination. The study provides a traceable analysis framework and operating-point hypotheses for independent evaluation; the small development cohort does not establish diagnostic utility or a reliable fusion or shape advantage.
\end{abstract}

\section{Introduction}
\label{sec:intro}
Single-marker measurements may fail to identify patients whose responses are more prominent in another marker. The experimental motivation is to combine PD-L1 and Siglec-15 measurements to reduce missed cancer cases. We examine whether that motivation is supported by patient-level paired response summaries, and whether the temporal allocation of an injection response adds information beyond its post-injection level. The empirical targets are improved discrimination and additional case detection at a comparable false-positive cost.

Three candidate changes must be evaluated separately. Selecting a retained-response window may change performance relative to an amplitude readout from the same target-related phase. Combining markers may improve on either single marker. Finally, an early area fraction may complement the combined retained-response score. Each comparison is required to identify the contribution of the corresponding change. We therefore retain a short, explicit comparison sequence and report its evaluable patient sets.

This exploratory pilot study has three linked objectives: to trace the selected curves to their original numerical sources; to compare retained response, marker fusion, and early shape on explicitly matched patient sets; and to examine the sensitivity of these comparisons to candidate timing and source aggregation. The procedure is fixed for the reported computation, but its development used the available cohort and is not prospective validation. The shape-evaluable patients form a nested subset of that same cohort. The results quantify preliminary discrimination and uncertainty and identify requirements for independent evaluation.

\subsection{Relationship to existing SPR research}
Huang et al.\ \cite{huang2021pdl1} developed a PD-L1 SPR system using antibody and aptamer recognition of the same target and evaluated human plasma samples from healthy donors and cancer patients. That dual-recognition strategy concerns one analyte. Their Siglec-15 work \cite{huang2021sig} uses magnetic alignment of plasmonic nanochains to improve angular response. These studies establish relevant sensing and interface foundations.

Wang et al.\ \cite{wang2025} subsequently reported a sensitized dual immune-checkpoint SPR biosensor involving PD-L1 and Siglec-15. Dual-marker detection therefore has direct precedent. The acquisition description associated with the present data identifies pSC4, whereas that published platform concerns CD-AuNP/viologen self-assembly. These descriptions do not establish that the present records are the original clinical dataset of that publication. Nor has equivalence of the clinical readout window, weighting rule, or analysis population been established. The target-phase amplitude comparator used here is therefore defined explicitly for this analysis and is not presented as an exact reproduction of the published method.

Dynamic SPR descriptors also have precedent. Grasmeier et al.\ \cite{grasmeier2023} studied dissociation properties in therapeutic monitoring of infliximab and anti-drug antibodies. Vollenberg et al.\ \cite{vollenberg2019} used the retained-to-end-association response ratio $R_{700}/R_{350}$ to characterize patient-derived anti-glycoprotein-V antibodies, and Metayer et al.\ \cite{metayer2025} used the area under normalized dissociation curves to characterize monoclonal-antibody glycosylation. The normalized retained-response ratio differs from the absolute retained response studied here, and dissociation-area descriptors differ from the early injection fraction. These applications motivate response-shape summaries but do not establish their incremental value for the present cancer-versus-healthy comparison.

Volpe et al.\ \cite{volpe2025} applied self-organizing maps to SPR trajectories in canine visceral leishmaniasis. Their study reports useful temporal information and comparison with univariate readouts. Its analytes, species, sampling, and assay differ from the present data, so its performance is not a numerical benchmark for this cohort. The contribution of the present study is a traceable, low-dimensional paired-marker analysis with explicit measurement and evaluation assumptions; it does not depend on claiming that marker fusion or temporal SPR descriptors are themselves new.

\section{Data selection and study population}
\label{sec:data}
\label{sec:rawaudit}
\subsection{Original exports and retained data}
The data comprise original numerical exports and a subsequent quality-selected set. The original exports contain 91 eight-column numerical files, each with four time--response column pairs. The selected set contains 122 two-column numerical files and an acquisition description. According to the investigators, manual acquisition introduced substantial error, and unusable data had already been processed or excluded before this analysis. We use this existing selected set without repeating the investigators' quality selection or restoring excluded sources. The available records permit an audit of which sources and rows were retained, but not a complete reconstruction of the reasons for every exclusion.

Numerical matching maps retained records to their original files, column pairs, and row intervals. Matching accommodates the retained export precision using absolute tolerances of $1.1\times10^{-5}$ for time and $5.1\times10^{-6}$ for response, with zero relative tolerance. Finite responses and timestamps are compared; partial transformations and nonunique same-value sources are recorded separately. These tolerances are matching rules, not biological quality thresholds. The complete mappings and input hashes are retained in the research archive; the ancillary code documents the downstream analysis and visualization.

The sources of the retained curves lie in original column pairs 1 and 3, corresponding to columns 1/2 and 5/6. The sole explicit text header assigns 670\,nm to these pairs and 785\,nm to pairs 2 and 4. Applying this wavelength ordering to other unlabelled files is a shared-format inference. The unused column pairs are distinguished from quality-selected exclusions. No wavelength is used as a substitute for a marker label.

\begin{table}[!ht]
\centering
\small
\caption{Source retention in the stable paired cohort (Group 2). Both markers refer to the same 20 patients (10 cancer and 10 healthy).}
\label{tab:source-retention}
\begin{tabular}{lrr}
\toprule
Quantity & PD-L1 & Siglec-15 \\
\midrule
Original column-pair slots & 40 & 40 \\
Retained files & 34 & 35 \\
Distinct retained sources & 34 & 34 \\
Not retained, with finite responses & 6 & 5 \\
Not retained, all responses missing & 0 & 1 \\
Patients with one retained source & 6 & 6 \\
Patients with two retained sources & 14 & 14 \\
\bottomrule
\end{tabular}
\par\smallskip
\begin{minipage}{0.95\linewidth}
\footnotesize A source is an original time--response column pair (columns 1--2 or 5--6). Different sources can be correlated and are not additional patients. The H2 Siglec-15 B/F files represent the same source, explaining the difference between file and source counts. Counts describe observed retention; finite responses alone do not establish suitability for analysis.
\end{minipage}
\end{table}

\subsection{Primary paired cohort and technical sources}
The primary exploratory cohort is group 2: study codes C1--C10 and H1--H10, representing 10 cancer and 10 healthy patients in the experimental design. Every code retains both PD-L1 and Siglec-15 measurements. The prepared files comprise 34 PD-L1 and 35 Siglec-15 curves. The original first/third column pairs provide 80 source slots: 68 are represented, 11 contain finite responses but were not retained, and one has entirely missing responses and was not retained.

The analyzed size is determined by these available records. No prospective sample-size calculation for these patient-level comparisons is documented in the available study records. Additional source curves, measured time points, and case--control score pairs do not increase the number of independent patients. No part of this cohort is represented as an untouched external test set.

The two retained H2 Siglec-15 files refer to the same original first column pair, up to export rounding. They therefore contribute one source, with the B-labelled export selected as the recorded canonical file. This yields 34 retained sources for each marker and 68 in total. Distinct source column pairs may still have correlated technical errors; neither the files nor the source pairs are independent patients. Patient--marker feature aggregation uses the available distinct sources and gives each patient equal weight.

\begin{figure}[!ht]
\centering
\includegraphics[width=\textwidth]{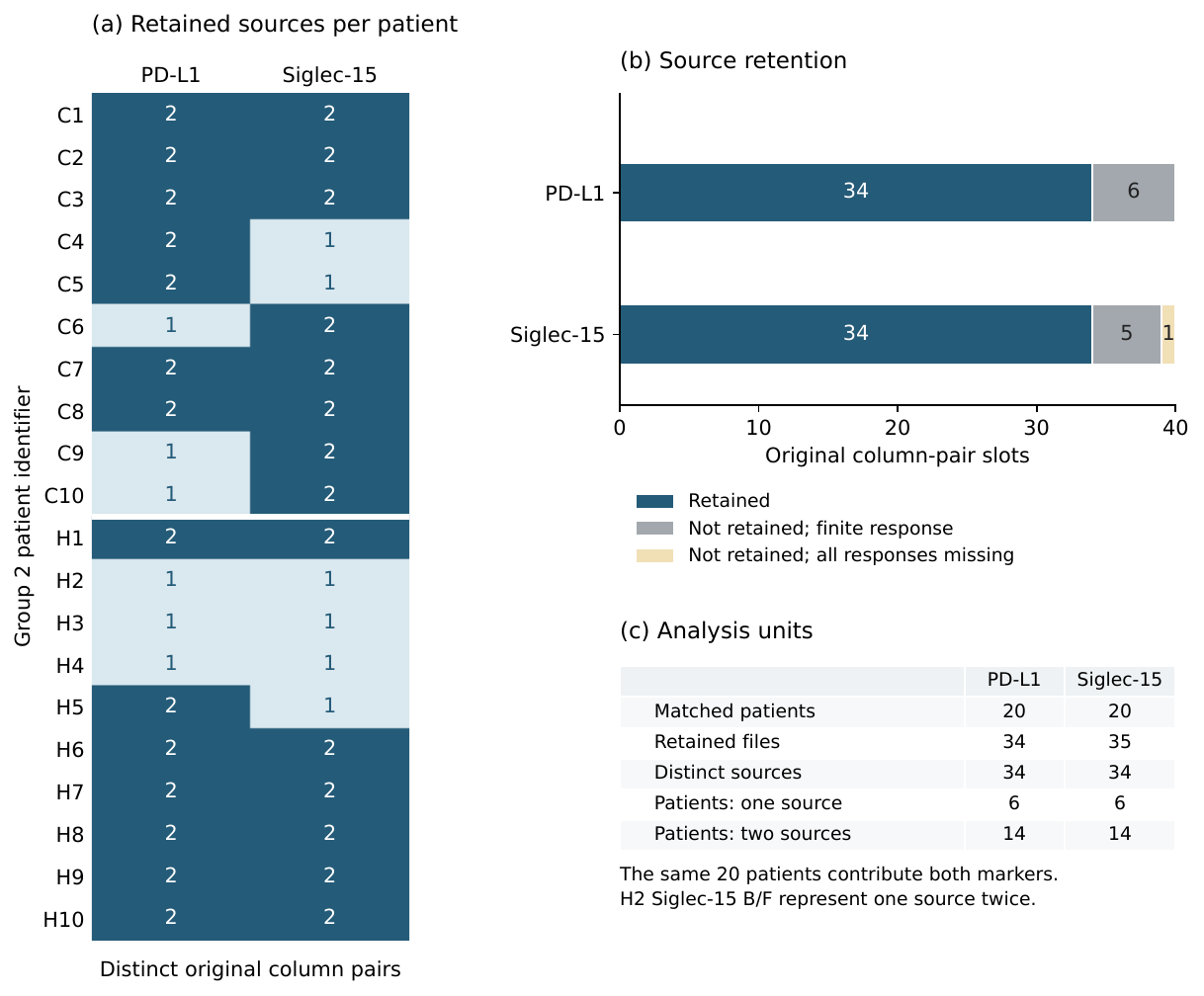}
\caption{Observed source coverage after the investigators' existing quality selection. The patient--marker counts refer to distinct original column pairs, not independent patients or assumed independent injections. Both markers remain represented for all 20 group-2 study patients. File and source counts differ because the two H2 Siglec-15 exports share one source. This figure does not depend on candidate event alignment.}
\label{fig:sourcecoverage}
\end{figure}

Selection also occurs within retained curves. For example, PD-L1 H6 retains the first 72 of 369 numerical rows in each source, while H10 retains the first 74 of 423. C9-F omits seven rows whose original responses are missing. Siglec-15 C5B omits seven finite-response rows; their specific selection rationale cannot be inferred from the numerical difference alone. Truncation and value processing are recorded separately. Appendix~\ref{sec:alignment-implementation} documents these distinctions; the row-level mapping tables are retained in the research archive.

Group 1 is not pooled into the primary cohort: its retained filenames include six cancer and three healthy codes, whereas the acquisition description states five cancer and three healthy samples. Group 3 contains five post-immunotherapy study codes, but no verified pretreatment pairing or treatment-response outcome is available. Its original exports also contain identical cross-marker content and a repeated time block. These groups are included in the source audit but do not provide independent validation cohorts for the present comparison.

\subsection{Observation model and historical relative windows}
Let $i$ index study patients, $m\in\{\mathrm{PDL1},\mathrm{SIG}\}$ index markers, and $r$ index distinct available technical sources. A source supplies ordered, possibly irregular times $t_{imrj}$ and responses $Y_{imrj}$. Write $G_i=1$ for cancer and $G_i=0$ for healthy controls. An explanatory measurement model is
\begin{equation}
Y_{imrj}=b_{imr}+f_{im}(t_{imrj})+d_{imr}(t_{imrj})+\varepsilon_{imrj},
\label{eq:obs}
\end{equation}
where $b_{imr}$ is an offset and $d_{imr}$ describes drift or systematic source departures. Temporal and between-source dependence are permitted. Zero-mean observation noise, when assumed in a mathematical result, does not by itself make a sample median or area ratio unbiased.

The inherited feature specification defines the relative windows in Table~\ref{tab:windows} after manual alignment and truncation of target-relevant segments. The selected numerical exports retain acquisition time, and the historical per-source manual alignment offsets are unavailable. File-level selection and target-phase alignment are therefore separate operations: matching original and selected exports recovers sources and rows, but does not establish the corresponding relative-time origins. For the present numerical comparisons, separately recorded candidate intervals localize visible response transitions, as described in Section~\ref{sec:conditional-method}. These reconstructed intervals are neither verified injection-valve event logs nor a recovery of the historical manual alignment. All timing-dependent results are conditional on this candidate reconstruction.

\begin{table}[!ht]
\centering\small
\caption{Historical relative-time definitions, applied after target-phase alignment.}
\label{tab:windows}
\begin{tabularx}{\textwidth}{@{}llL@{}}
\toprule
Window & Interval (min) & Interpretation\\
\midrule
$W_b$ & $[0,6)$ & Pre-target buffer baseline\\
$W_E$ & $[6,12]$ & First six minutes of injection\\
$W_I$ & $[6,16]$ & Ten-minute target injection\\
$W_R$ & $[16,20]$ & Early post-injection retained response\\
$W_L$ & $[20,24]$ & Later wash observation\\
\bottomrule
\end{tabularx}
\end{table}
\section{Scan-level features and quality requirements}
\label{sec:features}
The definitions in this section use target-relative time, with injection beginning at 6 min and ending at 16 min. They are not applied directly to unaligned acquisition timestamps. Section~\ref{sec:conditional-method} specifies the candidate-clock implementation, including the treatment of incompatible injection durations.

\subsection{Baseline and retained response}
Write $\mathcal I_{imr}(W)=\{j:t_{imrj}\in W\}$. For the recorded-point summary, set
\begin{align}
\widehat b_{imr}&=\Med\{Y_{imrj}:j\in\mathcal I_{imr}(W_b)\}, &
\widetilde Y_{imrj}&=Y_{imrj}-\widehat b_{imr},\label{eq:baseline}\\
\widehat R_{imr}&=\Med\{\widetilde Y_{imrj}:j\in\mathcal I_{imr}(W_R)\}.&&\label{eq:retention}
\end{align}
In the executed analysis, each median window must contain at least one recorded observation, and the acquisition interval must cover both endpoints of the requested window. Otherwise its summary is unavailable. The observed injection-peak comparator likewise requires at least one recorded point and acquisition coverage of its full window.

The descriptor is termed \emph{post-injection retained response}. A high value may reflect more signal at wash initiation, slower apparent loss, or both. It is not a direct estimate of molecular binding stability. Transport and rebinding can also affect apparent SPR kinetics \cite{carroll2016}.

An unweighted sample median gives each recorded point equal weight. With irregular acquisition, a densely sampled subinterval may therefore dominate. A time-weighted median or a median on a prespecified uniform-time grid targets a different, explicitly time-weighted summary. These are useful robustness analyses, but must not silently replace the recorded-point feature while retaining its reported results. The ideal time-median formula derived below applies to uniform time weighting. No formula equating a sample median to a midpoint response is assumed for an arbitrary acquisition grid.

\subsection{Areas and the early fraction}
Within $W_I$, start with one ordered grid containing all observed injection times and the boundaries 6, 12, and 16\,min. Insert a missing boundary only by interpolation between sufficiently close bracketing observations. Do not extrapolate an uncovered phase. A prespecified maximum gap and minimum coverage are required. Let $\ell_{imr}$ be the piecewise-linear interpolation of the corrected node values. Let $p_{imr}$ be the piecewise-linear interpolation of their positive parts, with $[x]_+=\max(x,0)$. Using the same grid for both areas gives
\begin{align}
\widehat A^E_{imr}&=\int_6^{12}\ell_{imr}(t)\,dt,\label{eq:earlyarea}\\
\widehat A^{E,+}_{imr}&=\int_6^{12}p_{imr}(t)\,dt,&
\widehat A^{I,+}_{imr}&=\int_6^{16}p_{imr}(t)\,dt,\label{eq:posareas}\\
\widehat E_{imr}&=\frac{\widehat A^{E,+}_{imr}}{\widehat A^{I,+}_{imr}}.&&\label{eq:earlyfrac}
\end{align}
The two integrals in \eqref{eq:posareas} are exactly trapezoidal integration of the clipped node values, using a fixed numerical convention. In general $p_{imr}(t)\ne[\ell_{imr}(t)]_+$ on a segment that crosses zero. Integrating the positive part after linear reconstruction is a distinct convention; it should be identified explicitly if used in a sensitivity analysis. Under adequate reconstruction, both approximate the continuous positive-response functional considered in the theory.

The numerator represents an \emph{area contribution} during the first six minutes of injection. It is not the fraction of binding events completed, and the first six minutes constitute 60\% of the injection duration, not its first half. The shared nonnegative reconstruction ensures $0\le\widehat E_{imr}\le1$ when the denominator is positive. Positivity alone, however, is insufficient for reliable interpretation.

A signed sensitivity descriptor is
\begin{equation}
\widehat E^{\mathrm{signed}}_{imr}
=\frac{\widehat A^E_{imr}}{\int_6^{16}\ell_{imr}(t)\,dt}.
\label{eq:signedfraction}
\end{equation}
The denominator must be separated from zero; the ratio need not lie in $[0,1]$.

\subsection{Additional comparators}
For $J_E=\mathcal I_{imr}(W_E)$ and $\overline t_E=|J_E|^{-1}\sum_{j\in J_E}t_{imrj}$, the ordinary least-squares slope is
\begin{equation}
\widehat S^E_{imr}=
\frac{\sum_{j\in J_E}(t_{imrj}-\overline t_E)\widetilde Y_{imrj}}
{\sum_{j\in J_E}(t_{imrj}-\overline t_E)^2}.
\label{eq:olsfeature}
\end{equation}
It requires at least two distinct times with adequate spread. The injection and late-window maxima are
\begin{equation}
\widehat M^I_{imr}=\max_{j\in\mathcal I_{imr}(W_I)}\widetilde Y_{imrj},\qquad
\widehat M^L_{imr}=\max_{j\in\mathcal I_{imr}(W_L)}\widetilde Y_{imrj}.
\label{eq:maximums}
\end{equation}
These descriptors distinguish alternative response summaries. The executed amplitude comparison uses the observed injection maximum; slopes and extra areas are not searched to select a better classifier. An ``overall maximum'' requires its exact search interval to be recorded before comparison. None is automatically identical to the published platform's amplitude readout.

\subsection{Weak signal, missingness, and quality flags}
The existing file- and row-level quality selection is retained throughout this analysis. For future acquisition, blank or independent quality-control measurements should determine minimum signal, minimum denominator area, admissible drift, boundary coverage, and maximum time-gap requirements before outcome-based comparisons. Their numerical values cannot be inferred from the available assay records. A shape feature failing its requirements is marked unavailable, rather than made finite by adding an arbitrary small denominator. The same rules apply to cases and controls. A shape failure need not imply a retention failure.

Positive clipping does not remove the weak-signal problem. For zero-mean stationary noise with positive-part expectation $a>0$, expected positive areas over the two windows are proportional to $6a$ and $10a$. Under a suitable concentration or ergodic averaging regime, their ratio can approach 0.60 without a target-specific association signal. This is a limiting explanation, not an equality between the expectation of a ratio and a ratio of expectations. The fraction's apparent amplitude cancellation under an ideal response model therefore does not establish reliability near the detection limit.

\section{Patient-level aggregation and dual-marker scoring}
\label{sec:score}
\subsection{Preserving technical-repeat information}
All features and quality flags are first retained at scan level. For a feature $x$ and its set of valid scans $\mathcal R^x_{im}$, the feature-wise aggregation is
\begin{equation}
\widehat x_{im}=\frac{1}{|\mathcal R^x_{im}|}\sum_{r\in\mathcal R^x_{im}}\widehat x_{imr},
\qquad |\mathcal R^x_{im}|\ge1.
\label{eq:aggregation}
\end{equation}
With exactly two scans, this equals the conventional two-observation median; with one scan it returns that scan. If additional scans become available, mean and median aggregation are no longer interchangeable. The rule must then remain prespecified. A verified repeated export of the same original source is counted once, using the canonical file recorded in the analysis manifest. Raw curves are not pooled before feature extraction. Patients with more repeats receive no greater weight in clinical performance estimates. Source availability is recorded, and dependence on source choice is examined in the aggregation sensitivity analysis.

\subsection{Two stages of training-based standardization}
For a training-patient set $\mathcal T$, let $\mu_{x,m,\mathcal T}$ and $s_{x,m,\mathcal T}$ be the mean and sample standard deviation of valid patient-level feature $x$ for marker $m$. Define
\begin{equation}
z_{x,m,\mathcal T}(\widehat x_{im})=
\frac{\widehat x_{im}-\mu_{x,m,\mathcal T}}{s_{x,m,\mathcal T}},
\qquad s_{x,m,\mathcal T}>0.
\label{eq:markerst}
\end{equation}
The dual-marker quantities are
\begin{align}
R_i^{\mathrm{sum}}&=z_{R,\mathrm{PDL1},\mathcal T}(\widehat R_{i,\mathrm{PDL1}})
+z_{R,\mathrm{SIG},\mathcal T}(\widehat R_{i,\mathrm{SIG}}),\label{eq:rsum}\\
R_i^{\max}&=\max\{z_{R,\mathrm{PDL1},\mathcal T}(\widehat R_{i,\mathrm{PDL1}}),
z_{R,\mathrm{SIG},\mathcal T}(\widehat R_{i,\mathrm{SIG}})\},\label{eq:rmax}\\
E_i^{\mathrm{sum}}&=z_{E,\mathrm{PDL1},\mathcal T}(\widehat E_{i,\mathrm{PDL1}})
+z_{E,\mathrm{SIG},\mathcal T}(\widehat E_{i,\mathrm{SIG}}).\label{eq:esum}
\end{align}
Training-only standardization of each combined block then gives
\begin{equation}
Z_{R,i}=z_{R^{\mathrm{sum}},\mathcal T}(R_i^{\mathrm{sum}}),\qquad
Z_{E,i}=z_{E^{\mathrm{sum}},\mathcal T}(E_i^{\mathrm{sum}}).
\label{eq:blockst}
\end{equation}
Both stages are fitted inside the training portion and frozen for held-out patients. Shape-marker moments use the corresponding valid training observations; combined shape moments use training patients with both shape measurements available. Each scaling stage requires at least two finite training values and a positive sample standard deviation. A zero-variance or incomplete leave-pair-out fold is reported as an error rather than assigned a fallback score. All reported folds satisfy these conditions.

\subsection{Primary score, candidate enhancement, and unavailable features}
The primary score is $Z_{R,i}$. The candidate enhanced score is
\begin{equation}
S_i(\lambda)=\lambda Z_{R,i}+(1-\lambda)Z_{E,i},
\qquad 0\le\lambda\le1.
\label{eq:composite}
\end{equation}
The historical exploration favored $\lambda\approx0.85$. The fixed candidate is consequently $S_i(0.85)=0.85Z_{R,i}+0.15Z_{E,i}$. This is a coefficient allocation, not a decomposition assigning 85\% of diagnostic information to retention. Any new weight optimization is a separate exploratory procedure requiring an inner training loop.

The primary incremental comparison is restricted to the common set with both shape descriptors and both retention descriptors available. Retention-only and enhanced scores are scaled and evaluated on that same set. Retention-only results for all 20 study patients are reported separately. An unavailable shape feature is not replaced by a small denominator or a fabricated observation. A full-cohort fallback rule that substitutes zero for an unavailable combined shape contribution would address a separate missing-feature policy and is not used for the primary shape comparison.

\section{Statistical analysis}
\label{sec:validation}
The comparisons below are exploratory analyses of a development cohort. Leave-pair-out assessment is used to examine training-only scaling while retaining all available patients for paired comparisons; a further small random train--test split is not used. No retrospective power calculation is used to reinterpret the observed estimates. Sample-size illustrations for a future fixed-rule validation study are presented separately in Section~\ref{sec:future-validation}.

\subsection{Recorded candidate alignment}
\label{sec:conditional-method}
Candidate intervals localize observed response transitions, not experimentally recorded valve switching. The annotation procedure first displayed the 40 patient--marker groups in a random order (NumPy seed 28747), with panel labels Q01--Q40 replacing patient and diagnosis text. Visual review identified the final substantial rise followed by a sustained falling or wash-like segment, using the stated assay sequence. The review usually examined the final 75 minutes of acquisition, extending to the complete curve or a magnified view when needed. Removing labels from these displays did not eliminate prior knowledge of the study; this was a project-informed reconstruction, not prospective blinded or independent adjudication.

The neighboring observed nodes before and after a visible rise defined the onset bracket; nodes on either side of the main falling transition defined the stop bracket. Gradual rises and multi-interval falls could receive additional extended bounds to retain visible ambiguity. Each paired scan was then inspected separately, producing 69 file-level entries; the initial 40-group reference timing was not imposed on both scans. Two files with unresolved transitions retained missing events, and the verified repeated export was counted once. The analysis thus uses 66 candidate-event records among 68 distinct sources. No patient's inclusion was reversed by this annotation procedure.

The recorded intervals and their midpoints are operational localization choices, not confidence intervals or guarantees that the physical switch occurred within them. They can reflect fluid transport, response dynamics, or sampling as well as switching. The annotations were not tuned against feature AUC or classification success and are held fixed in the analysis. Extended bounds are retained in the annotation record, but the reported common-shift sensitivity does not exhaust their uncertainty. The numerical consequences of this table are reproducible; the identity and precision of the physical events remain unverified.

For an onset bracket $[s_-,s_+]$ and stop bracket $[u_-,u_+]$, define
\begin{equation}
\mathcal J=[s_-,s_+]\cap[u_--10,u_+-10].
\label{eq:eventintersection}
\end{equation}
If $\mathcal J$ is nonempty, its midpoint defines $s$ and $u=s+10$. Otherwise the separate bracket midpoints define $s,u$ for retained response and the historical six-of-ten-minute shape fraction is unavailable. This rule does not classify an incompatible duration as an assay failure. Retention uses measured observations on $[u,u+4]$, with baseline observations on $[s-6,s)$. The peak comparator is the largest observed, baseline-corrected value during $[s,u]$; it does not require a reconstructable area fraction.

Areas use bracketing support, the common augmented grid, and the clipped-node trapezoidal convention of Section~\ref{sec:features}. A reconstruction gap exceeding 2.5\,min makes the area-based feature unavailable. This is a declared engineering guard reused from the recorded candidate analysis, not an independently established signal-quality cutoff. A positive area ensures mathematical definability only. The source-specific records retain all window counts and reasons for unavailable features. Additional implementation details appear in Appendix~\ref{sec:alignment-implementation}.

\subsection{Matched comparisons and apparent scores}
The all-patient comparisons use single-marker retained response, the standardized dual-marker retention sum, its max-rule sensitivity comparator, and an equivalently standardized dual-marker injection peak. Larger responses retain the case-positive direction fixed for this analysis; no AUC-dependent sign reversal is applied. The shape comparison uses the common subset with both markers' retention and early fraction available. Both retention-only and the fixed 85:15 score are recalculated on that same subset. This prevents a change in evaluable patients from masquerading as shape increment.

For apparent performance, both stages of scaling use the analyzed patients' sample moments. Apparent scores describe these patients and are not held-out predictions. The tie-aware sample AUC is
\begin{equation}
\widehat{\AUC}(T)=\frac{1}{n_1n_0}
\sum_{i:G_i=1}\sum_{j:G_j=0}
\left\{\mathbf1(T_i>T_j)+\tfrac12\mathbf1(T_i=T_j)\right\}.
\label{eq:empauc}
\end{equation}
Here $n_1,n_0$ count the same evaluable cases and controls for each paired comparison. Score availability is reported separately. A gain in correctly ordered case--control pairs is distinguished from the number of additional cases identified at a threshold.

For compact table notation, $R_\Sigma$ denotes the final standardized retention block $Z_R$, $P_\Sigma$ denotes the correspondingly standardized sum of marker injection peaks, and $S_{.85}$ denotes the fixed 85:15 score. When a table reports both the full and common-shape sets, their scaling populations are identified explicitly.

\subsection{Leave-pair-out internal assessment}
For every evaluable case $i$ and control $j$, hold out both patients and fit the scaling moments using all remaining patients. Score the held-out pair using that common fit, keeping the windows, marker combination, and 85:15 coefficient fixed. The internal summary is
\begin{equation}
\widehat{\AUC}_{\mathrm{LPO}}=
\frac{1}{n_1n_0}\sum_{i:G_i=1}\sum_{j:G_j=0}
\left\{\mathbf1(T_i^{(-i,-j)}>T_j^{(-i,-j)})
+\tfrac12\mathbf1(T_i^{(-i,-j)}=T_j^{(-i,-j)})\right\}.
\label{eq:lpoauc}
\end{equation}
All measurements belonging to a held-out patient are held out together. The two scores within a comparison share one training fit; scores from different fits are not pooled to construct an ordinary ROC curve. This executes training-only scaling without reserving an extremely small separate test set. Its overlapping training samples do not provide independent replications.

The current patients have already informed the feature definitions and candidate weight. Leave-pair-out assessment cannot remove that earlier selection or validate the event annotations. It is therefore conditional internal assessment of a fixed candidate procedure, not independent validation of the complete development process.

\subsection{Uncertainty and empirical false-positive budgets}
Paired, class-stratified patient bootstrap resampling is applied to the fixed apparent-score table. Each resampled patient carries both markers and both compared scores. The accompanying code uses 5,000 resamples with NumPy seed 20260910 and reports 95\% percentile intervals for AUC and paired differences. These intervals quantify variation of the fixed-score empirical discrimination statistic. They do not include refitting, historical feature selection, alignment uncertainty, or uncertainty in the leave-pair-out procedure; they are not validation intervals.

To examine missed-case patterns, each all-patient retention rule is evaluated under budgets of at most $b=0,1,2$ false positives among the ten healthy study patients. Write the ordered healthy scores as $T_{H,(1)}\ge\cdots\ge T_{H,(10)}$. For each rule, the implemented threshold and prediction are
\begin{equation}
c_b=T_{H,(b+1)},\qquad \widehat G_i(b)=\mathbf1\{T_i>c_b\}.
\label{eq:fpbudget}
\end{equation}
The cutoff is selected from the healthy scores after cohort scaling; cancer labels do not select the cutoff. Actual false positives can be fewer than the budget when tied control scores prevent an intermediate count. For each fusion rule we report newly identified and newly missed cancer cases relative to PD-L1 alone at the same budget.

These control-derived cutoffs are evaluated on the same cohort and are explicitly descriptive. They are not held-out sensitivity or specificity estimates. Ten healthy patients give empirical specificity increments of ten percentage points; a precise clinical claim at 95\% specificity is not supported. Reporting patient-level changes, indeterminate outcomes, and actual denominators follows the diagnostic-reporting rationale of Cohen et al.\ \cite{cohen2016}.

\subsection{Alignment and source-aggregation sensitivity}
The recorded candidate onset and stop times are translated together by $-1,0,+1$\,min, retaining the same selected source files. Features are re-extracted, including median-window membership and area support. No interpolated observation increases the effective sample count. The shape comparison is matched within each scenario, and scenario-specific sample counts are reported; a changing shape subset is not treated as the same target population.

Under a common translation by $h$, the feasible interval becomes $\mathcal J_h=\mathcal J+h$. Its emptiness and the candidate duration are unchanged. These scenarios therefore assess a common timing-origin displacement; they do not assess uncertainty in injection duration, independently displaced onset and stop events, or the strict ten-minute compatibility rule.

The ordinary available-source aggregate is also compared with deterministic first- and last-source alternatives among the distinct sources with a valid feature. These alternatives describe dependence on the technical source used; they do not select a source according to diagnosis or performance. Neither the shift ranges nor aggregation ranges are patient-sampling confidence intervals. The analyses do not optimize a new window or reverse an existing quality exclusion.
\section{Results}
\label{sec:conditional}
\label{sec:findings}
\subsection{Source coverage and evaluable features}
The completed source comparison establishes the selected cohort independently of any target-event assumptions (Figure~\ref{fig:sourcecoverage}). Of its 68 distinct sources, 66 have recorded candidate events. At zero shift, retained response and the observed injection peak are available from 33 PD-L1 and 33 Siglec-15 sources, supporting both markers for all 20 study patients. The early fraction is available from 20 PD-L1 and 14 Siglec-15 sources. The recorded reasons for unavailable shape features are ten-minute incompatibility for 32 distinct sources and missing candidate events for 2 sources. Both marker-level shape descriptors are available for 13 patients (7 cancer and 6 healthy).

For 27 of the 32 incompatible sources, the two intervals defining $\mathcal J$ are separated by less than 0.020\,min. This is a descriptive distance between recorded candidate brackets, not evidence of a biological failure or an estimate of clock error. No tolerance is added to the compatibility rule. The common-shape population consequently remains conditional on this strict candidate definition, whose uncertainty is not assessed by translating both events together.

\begin{figure}[!ht]
\centering
\includegraphics[width=\textwidth]{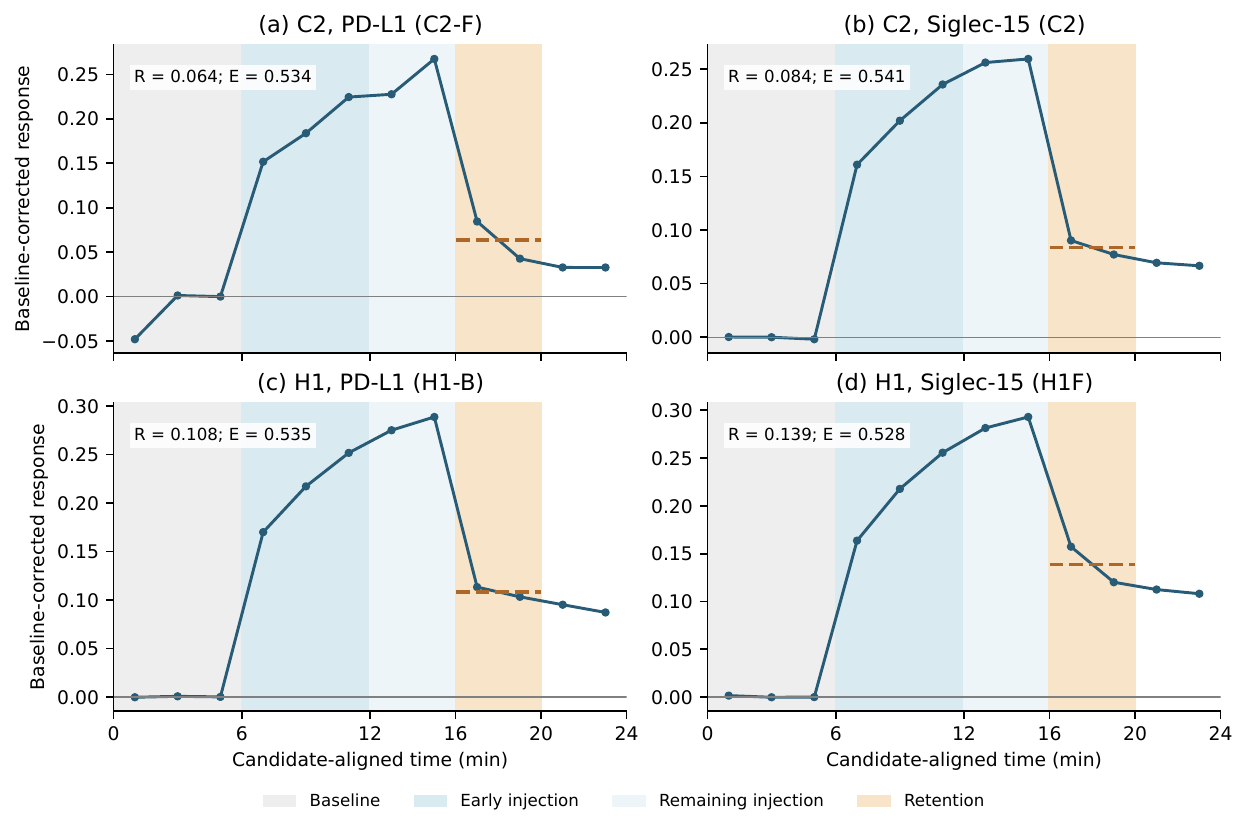}
\caption{Actual retained curves under recorded candidate alignment for C2 and H1. Dots are measured, baseline-corrected observations, joined for display; shaded regions identify the relative windows. Dashed segments mark the source-level retained median. These illustrations use candidate transition times and do not recover historical manual offsets. Source selection and the coordinates are exported by the accompanying Python script.}
\label{fig:alignedexamples}
\end{figure}

Figure~\ref{fig:alignedexamples} illustrates the candidate-relative windows on actual retained observations. The examples are the first naturally ordered shape-evaluable patient in each class, with the first source having a valid early fraction for each marker. They are chosen by a recorded illustration rule, without seeking the largest class difference. The complete patient-level feature table is reported in Appendix~\ref{sec:patient-table}.

\subsection{Amplitude and paired retention comparisons}
\begin{figure}[!ht]
\centering
\includegraphics[width=0.88\textwidth]{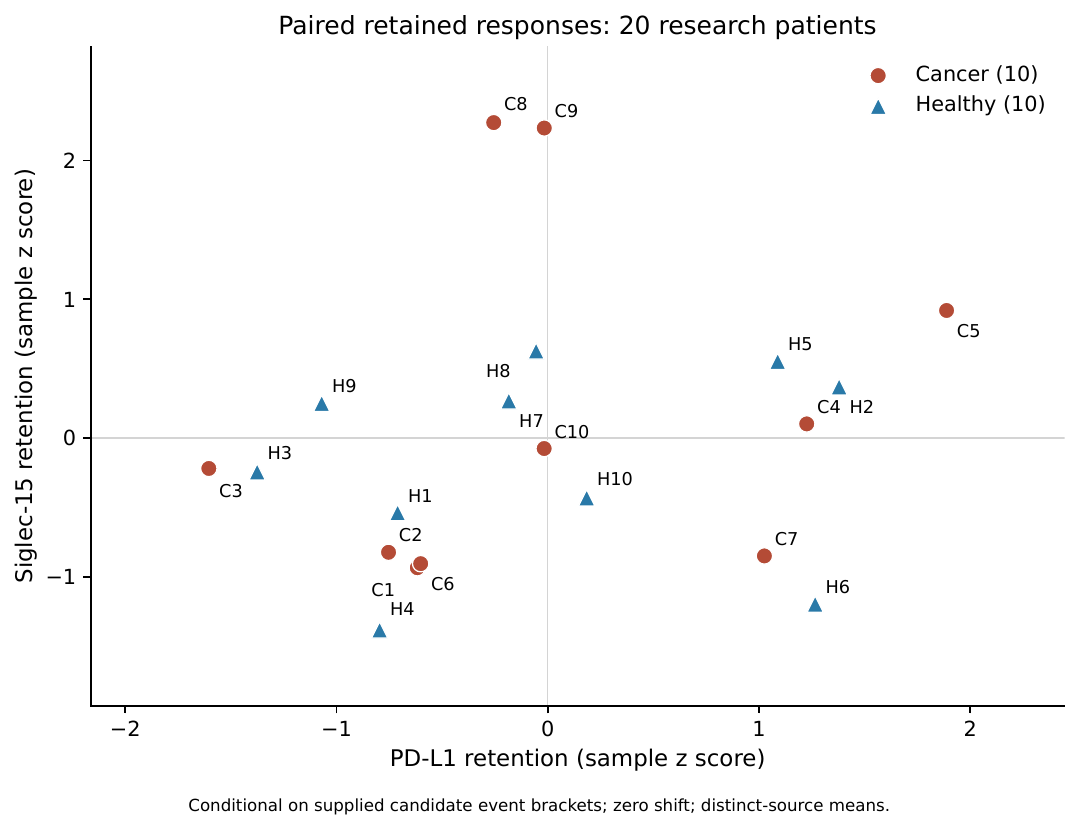}
\caption{Paired retained-response scores in the 20 study patients under unshifted candidate alignment. Marker values are standardized using this cohort. Each point is one patient; both marker measurements remain paired. The zero reference lines indicate cohort means, not clinical decision thresholds.}
\label{fig:pairedretention}
\end{figure}

The paired retention scatter shows substantial overlap between cases and controls (Figure~\ref{fig:pairedretention}). C8 and C9 have higher Siglec-15 responses than their PD-L1 responses would suggest on the respective marker scales, motivating inspection of their threshold classifications. This pattern is a within-study response observation, not evidence of universal biological mutual exclusion of PD-L1 and Siglec-15.

Table~\ref{tab:performance} reports apparent discrimination, fixed-score bootstrap intervals, and the completed leave-pair-out assessment. The dual retention sum has an apparent AUC of 0.570, compared with 0.520 for PD-L1, 0.530 for Siglec-15, and 0.460 for the dual injection peak. The corresponding paired differences are reported in Table~\ref{tab:increments}; their intervals include zero. The data under these candidate windows therefore do not establish a precise performance advantage.

\begin{table}[!ht]
\centering\small
\caption{Conditional patient-level discrimination under unshifted candidate alignment. Full denotes the same 20 patients; Shape denotes the same 13 patients with both shape markers available. Intervals use 5,000 class-stratified resamples of fixed apparent scores. LPO is leave-pair-out assessment with training-only scaling; it is not independent validation.}
\label{tab:performance}
\setlength{\tabcolsep}{4pt}
\begin{tabular}{llrrrr}
\toprule
Readout & Set & C/H & AUC & 95\% interval & LPO \\
\midrule
PD-L1 retention & Full & 10/10 & 0.520 & [0.260, 0.780] & 0.520 \\
Siglec-15 retention & Full & 10/10 & 0.530 & [0.250, 0.780] & 0.530 \\
Dual retention sum & Full & 10/10 & 0.570 & [0.300, 0.820] & 0.570 \\
Dual retention maximum & Full & 10/10 & 0.540 & [0.260, 0.800] & 0.540 \\
Dual injection peak & Full & 10/10 & 0.460 & [0.190, 0.720] & 0.460 \\
Dual retention sum & Shape & 7/6 & 0.619 & [0.286, 0.929] & 0.619 \\
Retention + shape (85:15) & Shape & 7/6 & 0.619 & [0.286, 0.929] & 0.619 \\
\bottomrule
\end{tabular}

\end{table}

For these data, all seven leave-pair-out AUCs equal their corresponding apparent AUCs. Training-only scaling changes multivariate score values but does not change the held-out pair-ordering totals. No window, weight, or orientation was reselected during internal assessment. These quantities remain conditional on the existing annotations and retrospective development choices.

\begin{figure}[!ht]
\centering
\includegraphics[width=\textwidth]{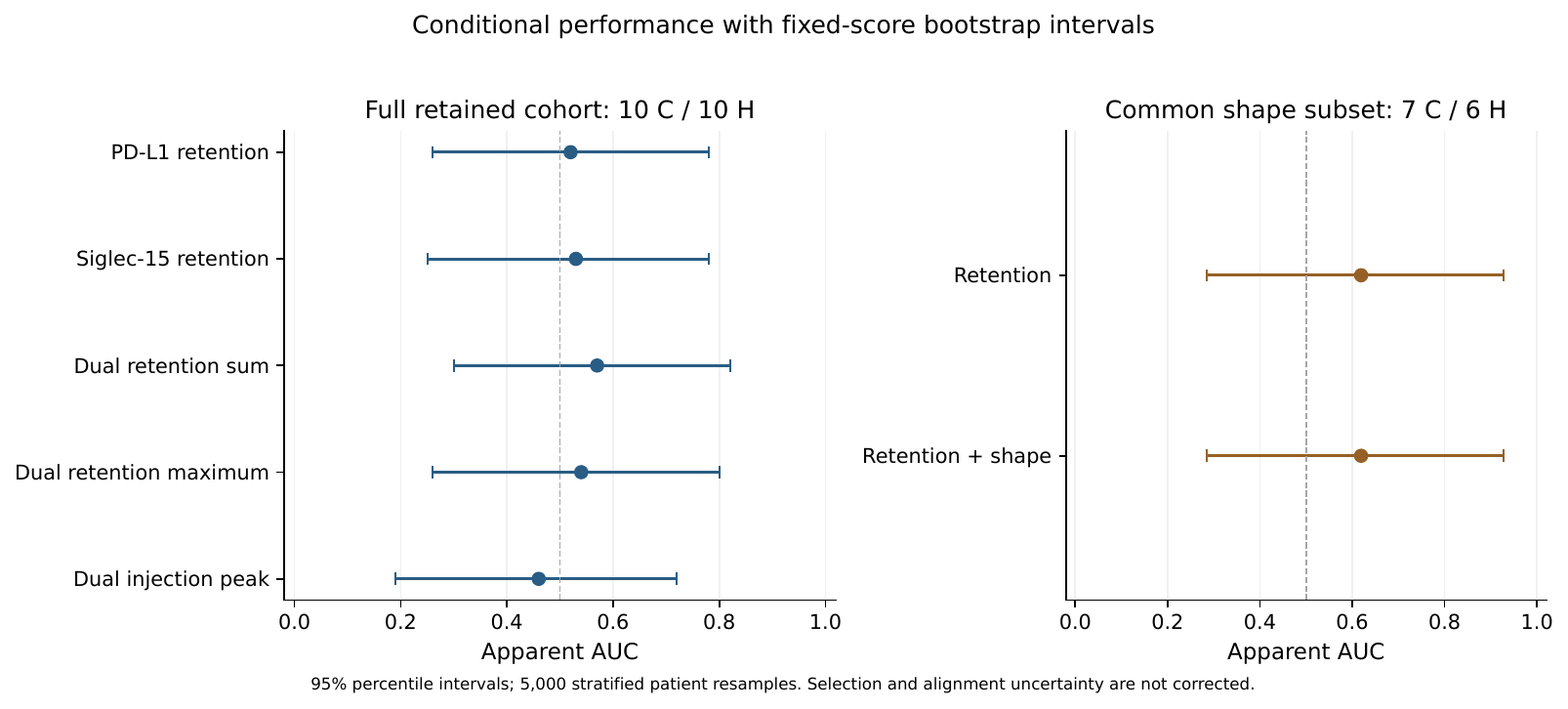}
\caption{Apparent AUCs conditional on the recorded candidate events. Error bars are percentile intervals from the paired, class-stratified fixed-score patient bootstrap. They omit scaling refitting, event uncertainty, and prior feature selection. The right-hand shape comparison uses the common 13-patient subset. Paired differences are reported separately in Table~\ref{tab:increments}.}
\label{fig:discrimination}
\end{figure}

\begin{table}[!ht]
\centering\small
\caption{Matched apparent AUC differences. Each difference and interval uses the same patients and bootstrap indices for both compared scores. These intervals do not quantify uncertainty of a fully refitted development procedure.}
\label{tab:increments}
\setlength{\tabcolsep}{5pt}
\begin{tabular}{lrrr}
\toprule
Paired comparison & C/H & $\Delta$AUC & 95\% interval \\
\midrule
Dual retention sum minus PD-L1 & 10/10 & +0.050 & [-0.150, +0.270] \\
Dual retention sum minus Siglec-15 & 10/10 & +0.040 & [-0.170, +0.250] \\
Dual retention sum minus dual peak & 10/10 & +0.110 & [-0.030, +0.290] \\
Retention + shape minus retention & 7/6 & +0.000 & [-0.095, +0.119] \\
\bottomrule
\end{tabular}

\end{table}

\subsection{Early-shape increment on the same patients}
On the common 7-case/6-control subset, both retention-only and the 85:15 score have AUC 0.619; the paired change is +0.000, with a fixed-score percentile interval of [-0.095, +0.119]. One case--control ordering improves and another worsens, yielding no net gain among the 42 pairs. The unchanged overall AUC does not mean that every patient's score or every pair ordering is unchanged.

The matched comparison isolates the addition of shape within one evaluable set. A full-cohort policy that substitutes zero for unavailable shape contributions would combine shape augmentation with a missing-feature policy and different scaling. Such a policy is not evaluated as evidence of shape increment here; all-patient retention results and common-subset shape results answer separate questions.

\subsection{Case detection at comparable false-positive budgets}
The empirical threshold table identifies actual patient transitions (Table~\ref{tab:fpbudgets}). With a budget of two healthy false positives, PD-L1 identifies 2 of ten cancer patients, Siglec-15 identifies 3, and both dual retention rules identify 4. Relative to PD-L1, the dual rules additionally identify C8 and C9 without losing another case. Siglec-15 alone identifies those cases but loses C4 relative to PD-L1 at this budget; the dual rules retain C4. The dual rules therefore have same-sample sensitivity 40\% and specificity 80\% at this operating point. At budgets zero and one, they match the Siglec-15 true-positive count rather than exceed it. These patient transitions suggest a local complementarity hypothesis; they do not establish improved clinical outcomes or independent diagnostic utility.

\begin{table}[!htbp]
\centering\small
\caption{Descriptive case detection under same-sample healthy-control false-positive budgets. All readouts use retained response; sum and maximum denote the two dual-marker rules. TP and FP count patients. Recovered and lost cases are relative to PD-L1 at the same budget. Each model's cutoff is selected from the healthy scores only and evaluated on this same cohort; these are not held-out clinical operating characteristics.}
\label{tab:fpbudgets}
\begin{tabular}{rlrrll}
\toprule
FP budget & Readout & TP & FP & Recovered & Lost \\
\midrule
0 & PD-L1 & 1/10 & 0/10 & -- & -- \\
0 & Siglec-15 & 3/10 & 0/10 & C8;C9 & -- \\
0 & sum & 3/10 & 0/10 & C8;C9 & -- \\
0 & maximum & 3/10 & 0/10 & C8;C9 & -- \\
1 & PD-L1 & 1/10 & 1/10 & -- & -- \\
1 & Siglec-15 & 3/10 & 1/10 & C8;C9 & -- \\
1 & sum & 3/10 & 1/10 & C8;C9 & -- \\
1 & maximum & 3/10 & 1/10 & C8;C9 & -- \\
2 & PD-L1 & 2/10 & 2/10 & -- & -- \\
2 & Siglec-15 & 3/10 & 2/10 & C8;C9 & C4 \\
2 & sum & 4/10 & 2/10 & C8;C9 & -- \\
2 & maximum & 4/10 & 2/10 & C8;C9 & -- \\
\bottomrule
\end{tabular}

\end{table}

\subsection{Timing and source-aggregation sensitivity}
\begin{figure}[!htbp]
\centering
\includegraphics[width=\textwidth]{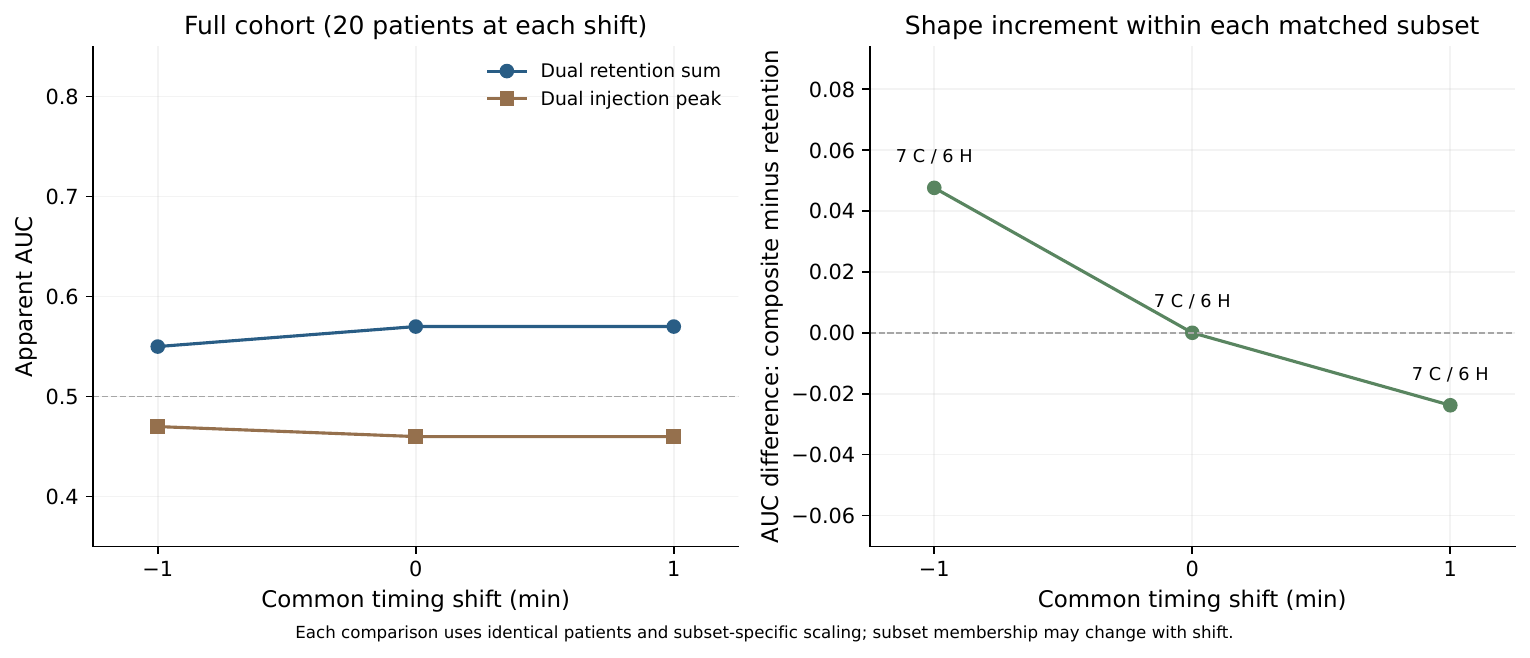}
\caption{Changes in apparent discrimination and matched shape increment after the declared common candidate-time translations. Retained source files and the scoring coefficient are fixed. Feature extraction, including actual median-window membership, is repeated. The common shape subset contains 7 cases and 6 controls in each displayed scenario.}
\label{fig:alignment-sensitivity}
\end{figure}

For common shifts of $-1,0,+1$\,min, the all-patient retention-sum AUCs are $0.550$, $0.570$, $0.570$. The shape AUC changes are $+0.048$, $+0.000$, $-0.024$ (Table~\ref{tab:sensitivity} and Figure~\ref{fig:alignment-sensitivity}). The same 13 patient identifiers remain shape-evaluable in all three scenarios. These changes correspond to net pair-ordering gains of two, zero, and minus one, respectively, among 42 comparisons; the positive and negative extremes compare scenarios two minutes apart. They indicate that a small shape increment is not stable across the declared scenarios, rather than a large deterioration established with precision. The comparisons share patients and are not independent replications.

\begin{table}[!htbp]
\centering\small
\caption{Sensitivity to translating all recorded candidate starts and stops by the stated number of minutes. The first two AUC columns use 10 cases and 10 controls; the final three use the matched shape subset listed in that row. $P_\Sigma$ is the dual injection-peak score. These are declared sensitivity scenarios, not confidence limits.}
\label{tab:sensitivity}
\setlength{\tabcolsep}{4.5pt}
\begin{tabular}{rrrrrrr}
\toprule
Shift & $R_\Sigma$ & $P_\Sigma$ & Shape C/H & $R_\Sigma$ (same) & $S_{.85}$ & $\Delta$ \\
\midrule
-1 & 0.550 & 0.470 & 7/6 & 0.548 & 0.595 & +0.048 \\
+0 & 0.570 & 0.460 & 7/6 & 0.619 & 0.619 & +0.000 \\
+1 & 0.570 & 0.460 & 7/6 & 0.619 & 0.595 & -0.024 \\
\bottomrule
\end{tabular}

\end{table}

Choosing the first or last available distinct source instead of the ordinary available-source aggregate gives retention-sum AUCs of 0.600 and 0.530, compared with 0.570 for averaging (Table~\ref{tab:aggregation}). These are deterministic source choices, not an optimization over sources. The result reinforces the need to report technical-source handling alongside patient-level scores.

\begin{table}[!ht]
\centering\small
\caption{Conditional apparent AUC under deterministic source-aggregation alternatives, using the same 20 patients and zero timing shift. First and last refer to lexicographic file order among distinct sources with a finite feature, not biological acquisition order.}
\label{tab:aggregation}
\begin{tabular}{lrrrrr}
\toprule
Aggregation & PD-L1 & Siglec-15 & $R_\Sigma$ & $R_{\max}$ & $P_\Sigma$ \\
\midrule
Mean of available & 0.520 & 0.530 & 0.570 & 0.540 & 0.460 \\
First available & 0.590 & 0.530 & 0.600 & 0.600 & 0.530 \\
Last available & 0.480 & 0.510 & 0.530 & 0.530 & 0.490 \\
\bottomrule
\end{tabular}

\end{table}
\section{Discussion}
\label{sec:discussion}
This exploratory pilot analysis establishes a traceable paired development cohort after the investigators' existing quality selection. Every group-2 patient retains both markers, although some patient--marker combinations contain only one retained source. The principal empirical findings are weak and imprecisely estimated discrimination under candidate alignment, no net shape increment at zero shift, and a local pattern of additional case detection at one empirical false-positive budget. These findings identify comparisons for further study without establishing a reliable diagnostic advantage.

Only 10 cases and 10 controls contribute to the full comparison; the shape comparison uses seven cases and six controls from the same cohort. One patient's classification changes the full-cohort sensitivity or specificity by ten percentage points. Reversing one case--control ordering changes the shape-subset AUC by $1/42\approx0.024$, and the 42 comparisons are not independent observations. The retention-sum interval [0.300, 0.820] illustrates the limited precision even under the restricted fixed-score bootstrap. Intervals containing zero for paired differences do not establish equivalence or the absence of useful information. Small sample size increases uncertainty but does not, by itself, explain the low AUC; a larger cohort could yield either a higher or a lower estimate. Repeated technical sources and resampling cannot supply additional independent patients.

The empirical comparison sequence distinguishes a change in readout from a second-marker contribution and an additional shape contribution. These questions remain relevant even if one proposed enhancement is small or inconsistent. A positive change in AUC need not correspond to fewer missed patients at a particular threshold, while a max rule can exchange additional case detection for additional healthy false positives. Actual patient transitions at comparable false-positive budgets therefore accompany the discrimination summaries.

The numerical analyses are conditional on recorded candidate response transitions. They quantify the output of an explicitly specified computation, including internal scaling assessment and timing sensitivity. Historical per-scan manual offsets and experimentally recorded switch times are unavailable, so the analysis does not establish faithful reproduction of a historical alignment or its performance. The present estimates characterize this candidate workflow; they cannot establish an intrinsic performance limit of the retained data or the SPR platform.

The mathematical development explains both interpretability and limitations. A positive area fraction cancels a common multiplicative response scale, but generally remains affected by concentration-dependent apparent rate, baseline error, noise, and sampling. Retention combines the amount of response present at wash initiation with subsequent response loss; it does not identify affinity or a unique decay rate. The supported-reconstruction bounds connect observation error, baseline error, alignment displacement, and maximum retained-grid gap under stated assumptions. The denominator restriction shows why a mathematically finite early fraction can still be unreliable in weak-response records. These bounds do not estimate the actual error components or establish that drift or noise caused the observed discrimination. Controlled simulations with known trajectories and experimental reference measurements would be needed to assess their practical sharpness and distinguish candidate mechanisms; those studies have not been performed here.

The study has a small case-versus-healthy cohort and cannot establish performance in a symptomatic differential-diagnosis population. Its uncertainty intervals have a deliberately limited fixed-score interpretation. Existing quality exclusions define the analyzed records; reconstructing their numerical consequences does not establish a prospective, outcome-blinded quality protocol. Diagnostic reference details, specimen form, and acquisition batch allocation should accompany any eventual clinical report. Group-3 treatment status without paired baseline measurements and response outcomes is insufficient for treatment-efficacy prediction.

\subsection{Planning an independent evaluation}
\label{sec:future-validation}
The immediate priority is to verify the target-stage event meaning and its correspondence to the intended experimental readout. If historical manual offsets cannot be recovered, the candidate workflow should be evaluated as a separately defined procedure. The existing 20 patients remain development data; a random subdivision cannot make them independent of the preceding feature and weight choices. Before new validation measurements, freeze the acquisition protocol, event-recording rule, feature definitions, source aggregation, scaling constants, coefficients, handling of indeterminate results, and cutoff, consistent with Pepe et al.\ \cite{pepe2008}.

The validation objective should then determine the sample size. Estimating sensitivity and specificity at a frozen threshold requires adequate numbers of cases and controls. Demonstrating an increment over a single marker requires a paired design based on a prespecified clinically relevant difference, the dependence between scores, and, for operating-point comparisons, discordant classifications. The small observed increments in this pilot should inform several planning scenarios rather than be treated as known population effects.

To illustrate the scale of a precision calculation, suppose a fixed rule has sensitivity $p_C=0.80$ and specificity $p_H=0.90$. These are hypothetical design inputs, not estimates or predictions of this SPR procedure. Inverting the usual marginal normal-approximation interval for a proportion gives
\begin{equation}
n_g^{\rm ill}(d)=\left\lceil\frac{1.96^2p_g(1-p_g)}{d^2}\right\rceil,
\qquad g\in\{C,H\},
\label{eq:illustrative-size}
\end{equation}
where $d$ is the target half-width of a two-sided 95\% interval and the denominators are evaluable cases and evaluable controls, respectively. The normal approximation and alternatives for small binomial samples are described in the NIST/SEMATECH handbook \cite{nistbinomial}. Table~\ref{tab:planning} is generated by a separate Python script that reads no patient data.

\begin{table}[!ht]
\centering\small
\caption{Hypothetical marginal precision planning for an independent fixed-rule validation study, assuming sensitivity 0.80 and specificity 0.90. Counts are normal-approximation illustrations, not observed outcomes or finalized recruitment targets.}
\label{tab:planning}
\begin{tabular}{lrr}
\toprule
Target 95\% interval half-width & Evaluable cases & Evaluable controls \\
\midrule
0.10 & 62 & 35 \\
0.05 & 246 & 139 \\
\bottomrule
\end{tabular}

\end{table}

These separate marginal calculations do not guarantee realized interval widths or simultaneous 95\% coverage, and they do not establish power for a paired AUC increment. Formal planning should use the chosen interval method or simulation over plausible performance values; the approximation is particularly crude for the 35-control example. The counts exclude development participants and make no allowance for unavailable or indeterminate results. Recruitment should accommodate prespecified class-specific evaluability assumptions, and the proportion and reasons for indeterminate measurements should be reported rather than silently discarded. The current 13/20 shape availability is conditional on the candidate definition and should not be assumed to be a stable future evaluability rate.

Validation cases and reference-negative participants should represent the intended clinical setting; healthy-only comparisons cannot establish performance among symptomatic patients with alternative diagnoses. Participants should be distributed across days, chips, and run positions to permit assessment of technical variation. Where mechanism is the objective, matched reference surfaces, buffer controls, concentration and flow-rate experiments, and an orthogonal assay measuring the same analyte form are needed. Independent patient evaluation and controlled measurement studies address different uncertainties.

\paragraph{Data and code availability.}
The ancillary files contain Python source for feature extraction, paired statistical comparisons, provenance visualization, and figure generation, together with an input-contract description and dependency versions. The ideal-trajectory illustration and hypothetical precision-planning scripts run without clinical inputs. Regenerating empirical results requires the original and quality-selected exports, the scan-specific candidate-event table, and the numerical source mapping. Those inputs are retained in the research archive and are not distributed with this preprint. The patient-level derived feature table is reported in Appendix~\ref{sec:patient-table}. The public source-code supplement therefore documents the computation but does not by itself enable independent regeneration of the empirical results. The LaTeX source includes all tables and figure assets required to compile the manuscript.

\paragraph{Generative AI assistance.}
ChatGPT was used to assist with analysis-code development, checking analytical derivations, and drafting and revising the manuscript. Such assistance does not provide independent experimental adjudication or clinical validation. Responsibility for the scientific claims and the submitted text remains with the author(s).

\clearpage
\appendix
\section{Analytical properties and their interpretation}
\label{sec:analytical-properties}

The results below characterize the proposed functionals under explicit mathematical assumptions. They do not establish diagnostic accuracy or identify a molecular binding mechanism from a clinical sensorgram. Throughout this section, $f$ denotes a continuous, baseline-corrected response; patient, marker, and scan indices are suppressed. The empirical retention and early-fraction features remain denoted by $\widehat R$ and $\widehat E$, respectively. Proofs are given in Appendix~\ref{sec:analytical-proofs} or immediately after the relevant statement.

\subsection{Association amplitude, slopes, and early-area fractions}

Consider the ideal association model
\begin{equation}
 f(t)=A\{1-\exp[-k(t-6)]\},\qquad 6\leq t\leq16,
 \quad A>0,\quad k>0.
 \label{eq:th-association}
\end{equation}
Here $A$ is the limiting response amplitude and $k$ is an apparent association-rate parameter. These quantities need not have separate biological interpretations in a heterogeneous clinical sample.

\begin{proposition}[Initial, secant, and least-squares slopes]
\label{th:slopes}
Under \eqref{eq:th-association}, the initial derivative and the early-window secant slope are
\begin{equation}
 f'(6)=Ak,\qquad
 S_{\mathrm{sec}}=\frac{f(12)-f(6)}{6}
                 =\frac{A(1-e^{-6k})}{6}.
 \label{eq:th-slopes}
\end{equation}
Both are strictly increasing in either $A$ or $k$ when the other parameter is fixed. For distinct observation offsets $u_1,\ldots,u_q\in[0,6]$, $q\geq2$, the ordinary least-squares slope with an intercept is instead
\begin{equation}
 S_{\mathrm{OLS}}(A,k)
 =-A\frac{\sum_{j=1}^{q}(u_j-\bar u)e^{-ku_j}}
           {\sum_{j=1}^{q}(u_j-\bar u)^2},
 \qquad \bar u=q^{-1}\sum_{j=1}^{q}u_j.
 \label{eq:th-ols-ideal}
\end{equation}
It is positive and scales linearly with $A$, but it need not be monotone in $k$.
\end{proposition}

The secant and OLS slopes are different functionals. In particular, the monotonicity of $S_{\mathrm{sec}}$ cannot be used as a monotonicity claim for the implemented OLS feature. Both slope descriptors retain an amplitude contribution.

For a continuous response on the injection interval, write
\begin{equation}
 N(f)=\int_6^{12}[f(t)]_+\,dt,\qquad
 D(f)=\int_6^{16}[f(t)]_+\,dt,\qquad
 E(f)=\frac{N(f)}{D(f)},\quad D(f)>0,
 \label{eq:th-continuous-fraction}
\end{equation}
where $[x]_+=\max(x,0)$. These are continuous-time functionals. Their numerical implementation requires the common boundary and reconstruction convention specified for the observed features.

\begin{proposition}[Area, amplitude invariance, and strict shape monotonicity]
\label{th:early-fraction}
Under \eqref{eq:th-association}, for $0<a\leq10$,
\begin{equation}
 \int_6^{6+a} f(t)\,dt
 = A Q_a(k),\qquad
 Q_a(k)=a-\frac{1-e^{-ak}}{k}.
 \label{eq:th-area-formula}
\end{equation}
Since $f\geq0$, the positive-part and signed area fractions coincide, and
\begin{equation}
 E(f)=E(k)
 =\frac{Q_6(k)}{Q_{10}(k)}
 =\frac{6k-1+e^{-6k}}{10k-1+e^{-10k}}.
 \label{eq:th-early-formula}
\end{equation}
The function $E(k)$ is strictly increasing for $k>0$, is independent of $A$, and satisfies
\begin{equation}
 \lim_{k\downarrow0}E(k)=\frac{9}{25},\qquad
 \lim_{k\to\infty}E(k)=\frac35,\qquad
 \frac{9}{25}<E(k)<\frac35\quad(k>0).
 \label{eq:th-early-range}
\end{equation}
More generally, $E(cf)=E(f)$ for every $c>0$ whenever $D(f)>0$, without requiring the exponential model.
\end{proposition}

The early fraction is the share of positive response area accumulated during the first six of ten injection minutes. It is not the fraction of molecules bound or the fraction of binding completed. Its amplitude invariance concerns multiplication of the entire curve by a positive constant. It does not imply invariance to a baseline error, additive noise, or a concentration-induced change in shape. Even under the elementary one-to-one association model, the apparent rate has the form $k=k_{\mathrm{on}}C+k_{\mathrm{off}}$ at fixed analyte concentration $C$; cancelling $A$ therefore does not make $E(k)$ concentration independent. The interval in \eqref{eq:th-early-range} is an ideal-model result, not a universal acceptance interval for observed curves. In particular, a value near $0.60$ alone does not establish rapid binding when the total response is weak.

\subsection{Wash response and the meaning of retention}

Consider the ideal wash model
\begin{equation}
 f(t)=R_{16}e^{-k_{\mathrm{off}}(t-16)},\qquad
 16\leq t\leq20,\quad R_{16}>0,\quad k_{\mathrm{off}}>0.
 \label{eq:th-wash}
\end{equation}
The parameter $R_{16}$ is the response at the start of wash, and $k_{\mathrm{off}}$ is an apparent decay rate.

\begin{proposition}[Continuous and sampled retention summaries]
\label{th:retention}
Under \eqref{eq:th-wash}, the time average and the median with respect to uniform time on $[16,20]$ are
\begin{equation}
 \mathcal R_{\mathrm{mean}}
 =\frac14\int_{16}^{20}f(t)\,dt
 =R_{16}\frac{1-e^{-4k_{\mathrm{off}}}}{4k_{\mathrm{off}}},
 \qquad
 \mathcal R_{\mathrm{time\text{-}med}}
 =R_{16}e^{-2k_{\mathrm{off}}}.
 \label{eq:th-retention-functionals}
\end{equation}
Both are strictly increasing in $R_{16}$ and strictly decreasing in $k_{\mathrm{off}}$ when the other parameter is held fixed.

For a fixed ordered sampling grid $16\leq t_1<\cdots<t_q\leq20$, the ordinary sample median of the noiseless values is
\begin{equation}
 \Med\{f(t_1),\ldots,f(t_q)\}
 =\begin{cases}
 R_{16}e^{-k_{\mathrm{off}}(t_{h+1}-16)},&q=2h+1,\\[2mm]
 \dfrac{R_{16}}{2}\{e^{-k_{\mathrm{off}}(t_h-16)}
             +e^{-k_{\mathrm{off}}(t_{h+1}-16)}\},&q=2h.
 \end{cases}
 \label{eq:th-sample-retention}
\end{equation}
It is increasing in $R_{16}$ and nonincreasing in $k_{\mathrm{off}}$, with strict decrease if at least one time entering the median is later than $16$.
\end{proposition}

The empirical $\widehat R$ uses a sample median and generally differs from the uniform-time median, especially with irregular sampling. For even sample sizes, the usual sample median also averages two response values; it need not equal the response evaluated at the average of their times. These distinctions should be retained when comparing recordings with different sampling patterns.

All the retention summaries in Proposition~\ref{th:retention} combine the initial wash amplitude with subsequent decay. For example, every pair $(R_{16},k_{\mathrm{off}})$ satisfying $R_{16}=r\exp(2k_{\mathrm{off}})$ produces the same positive time median $r$. Thus a larger retention value alone does not identify slower dissociation, greater affinity, or improved molecular stability. Retention is interpreted here as the amount of response remaining in a specified wash window.

\subsection{Technical repeats and marker combination}

\begin{proposition}[Two-scan averaging with correlated errors]
\label{th:replicate-variance}
Conditionally on a patient--marker feature value $\theta$, suppose
$X_B=\theta+\eta_B$ and $X_F=\theta+\eta_F$, where the errors have zero conditional means, positive conditional variances $\sigma_B^2$ and $\sigma_F^2$, and conditional correlation $\rho$. With the usual sample-median convention for two observations,
\begin{equation}
 \Med\{X_B,X_F\}=\frac{X_B+X_F}{2},\qquad
 \Var\left(\frac{X_B+X_F}{2}\,\middle|\,\theta\right)
 =\frac{\sigma_B^2+\sigma_F^2+2\rho\sigma_B\sigma_F}{4}.
 \label{eq:th-replicate-variance}
\end{equation}
If $\sigma_B^2=\sigma_F^2=\sigma^2$, this variance is $\sigma^2(1+\rho)/2$. The value $\sigma^2/2$ holds for uncorrelated errors, including the independent-error special case.
\end{proposition}

This calculation concerns already extracted scalar features; it does not require averaging the raw sensorgrams. A common baseline or batch error can induce correlation between scans and limit the benefit of averaging. With unequal error variances, equal-weight averaging need not be more precise than the better single scan. Moreover, a zero-mean observation error does not by itself imply a zero-mean error for a sample median or an area ratio. The feature-level assumptions in Proposition~\ref{th:replicate-variance} must therefore be assessed rather than inferred from an additive observation model. Technical repeats continue to belong to one patient for validation and uncertainty estimation.

\begin{proposition}[Sufficient support conditions for simple fusion rules]
\label{th:fusion-support}
Let $(x_1,x_2)$ be two prespecified marker scores on a fixed scale, with larger values oriented towards the case group.
\begin{enumerate}
 \item If every control has $x_1\leq a$ and $x_2\leq a$, while every case has at least one coordinate at least $b>a$, then the rule $\max(x_1,x_2)>c$ separates the two groups for every $a<c<b$.
 \item If every control has $x_1+x_2\leq a$, while every case has $x_1+x_2\geq b>a$, then the rule $x_1+x_2>c$ separates the two groups for every $a<c<b$.
\end{enumerate}
\end{proposition}

These are illustrative sufficient conditions on nonoverlapping score supports. They do not establish that the conditions hold in the SPR data, that either fusion rule improves AUC, or that one rule dominates the other. In particular, at the same numerical threshold a max rule can increase the control-positive probability. A fair empirical comparison must estimate each rule's threshold using training patients and compare sensitivity at a common specificity. The choice of sum as a primary candidate and max as a sensitivity analysis is a choice fixed for the present analysis, rather than a consequence of Proposition~\ref{th:fusion-support}.

\subsection{Numerical perturbation and weak-response limitations}

\begin{lemma}[Integral and positive-area-ratio perturbation bounds]
\label{th:reconstruction}
Let $f$ and $g$ be continuous on $[6,16]$, and let
$\delta=\sup_{6\leq t\leq16}|g(t)-f(t)|$.
For every interval $W\subseteq[6,16]$,
\begin{equation}
 \left|\int_W(g-f)\,dt\right|\leq |W|\delta,
 \qquad
 \left|\int_W[g]_+\,dt-\int_W[f]_+\,dt\right|
 \leq |W|\delta.
 \label{eq:th-integral-bound}
\end{equation}
The corresponding average errors are at most $\delta$.

Use the definitions in \eqref{eq:th-continuous-fraction}, and set
$D=D(f)$, $E=E(f)$, $\Delta N=N(g)-N(f)$, and $\Delta D=D(g)-D(f)$.
If $D\geq d_0>0$ and $10\delta<d_0$, then $D(g)>0$ and
\begin{equation}
 E(g)-E(f)
 =\frac{\Delta N-E\Delta D}{D+\Delta D},
 \qquad
 |E(g)-E(f)|
 \leq\frac{\{6(1-E)+4E\}\delta}{d_0-10\delta}
 \leq\frac{6\delta}{d_0-10\delta}.
 \label{eq:th-ratio-bound}
\end{equation}
Its first-order expansion in the area perturbations has the exact remainder representation
\begin{equation}
 \begin{aligned}
 E(g)-E(f)&=\frac{\Delta N-E\Delta D}{D}+r,\qquad
 r=-\frac{(\Delta N-E\Delta D)\Delta D}{D(D+\Delta D)},\\
 |r|&\leq\frac{60\delta^2}{d_0(d_0-10\delta)}.
 \end{aligned}
 \label{eq:th-ratio-remainder}
\end{equation}
Consequently, uniform convergence of reconstructions implies convergence of these areas, averages, and the fraction whenever the limiting denominator is positive.
\end{lemma}

The denominator condition is substantive. Positive-part clipping makes the area functional Lipschitz but does not make a near-zero-denominator ratio stable. The expansion in \eqref{eq:th-ratio-remainder} is in the area perturbations; no differentiability of $x\mapsto[x]_+$ at zero is assumed. Nor does interpolation of increasingly many noisy observations, by itself, establish the uniform convergence required by the lemma. Integral accuracy, sampling gaps, baseline uncertainty, and signal adequacy remain separate quality-control considerations. Trapezoidal integration of clipped sample values and integration after clipping a linearly reconstructed signed curve also differ on intervals containing a sign change; one fixed implementation must be used throughout.

\begin{lemma}[Fixed-design stability of slopes and sample medians]
\label{th:fixed-design}
Let $t_1,\ldots,t_q$ be fixed times with $q\geq2$ and
$S_{tt}=\sum_{j=1}^{q}(t_j-\bar t)^2>0$.
For two response vectors $y,z\in\R^q$ satisfying
$\max_j|y_j-z_j|\leq\delta$, their OLS slopes with an intercept and their ordinary sample medians satisfy
\begin{equation}
 |S_{\mathrm{OLS}}(y)-S_{\mathrm{OLS}}(z)|
 \leq\delta\frac{\sum_{j=1}^{q}|t_j-\bar t|}{S_{tt}}
 \leq\delta\sqrt{\frac{q}{S_{tt}}},
 \qquad
 |\Med(y)-\Med(z)|\leq\delta.
 \label{eq:th-fixed-design-bound}
\end{equation}
The OLS slope is unchanged by adding a common constant to all responses.
\end{lemma}

Thus these summaries are continuous in the observed values on a fixed nondegenerate design. The OLS bound also exhibits the effect of limited temporal spread through $S_{tt}$. It does not justify treating interpolated points as independent observations, or replacing a sampling-weighted OLS slope by a time-weighted slope without changing the estimand. A constant baseline shift cancels from an intercept-based OLS slope; the same invariance does not hold for retention levels or positive-area fractions.

\subsection{Propagation of baseline, sampling, and alignment error}
\label{sec:propagation-errors}

The preceding continuity bounds can be linked directly to the retained
observations and the implemented clipped-node trapezoidal rule. The first
result compares two versions of the same retained acquisition design; it
does not assume that new observations become available.

\begin{proposition}[Baseline correction on a fixed retained design]
\label{th:baseline-grid}
Let $y_j,z_j$ be two response vectors on the same time grid. Suppose that
the nonempty baseline and retention index sets are unchanged and that
$\max_j|y_j-z_j|\leq\epsilon$ over all baseline, retention, and
area-reconstruction observations. Define each baseline by its ordinary
sample median and subtract it before extracting features. Then
\begin{equation}
 |\widehat b_y-\widehat b_z|\leq\epsilon,
 \qquad
 \max_j|\widetilde y_j-\widetilde z_j|\leq2\epsilon,
 \qquad
 |\widehat R_y-\widehat R_z|\leq2\epsilon.
 \label{eq:baseline-grid-error}
\end{equation}
Construct the positive reconstructions $p_y,p_z$ on the same supported
grid, inserting the boundaries $6,12,16$ by linear interpolation before
clipping their node values. Then
$\|p_y-p_z\|_\infty\leq2\epsilon$ on $[6,16]$.
Write $D_y=\int_6^{16}p_y(t)\,dt$ and
$E_y=D_y^{-1}\int_6^{12}p_y(t)\,dt$, and define $D_z,E_z$ similarly.
If $D_y\geq d_0>20\epsilon$, then
\begin{equation}
 D_z\geq d_0-20\epsilon>0,
 \qquad
 |E_z-E_y|
 \leq
 \frac{2\epsilon\{6(1-E_y)+4E_y\}}{d_0-20\epsilon}
 \leq\frac{12\epsilon}{d_0-20\epsilon}.
 \label{eq:baseline-grid-ratio}
\end{equation}
The sampling grid need not be uniform.
\end{proposition}

\begin{proof}
The sample-median bound in Lemma~\ref{th:fixed-design} gives the first
inequality in \eqref{eq:baseline-grid-error}; its proof also applies to a
single observation. Subtracting the two estimated baselines gives the
second inequality, and applying the median bound to the corrected
retention values gives the third. Linear interpolation forms convex
combinations, and positive-part clipping is $1$-Lipschitz. Consequently,
boundary interpolation, node clipping, and the final interpolation each
preserve the $2\epsilon$ bound. The early and remaining injection areas
therefore differ by at most $12\epsilon$ and $8\epsilon$, respectively.
The same nested-area identity used in Lemma~\ref{th:reconstruction} gives
\eqref{eq:baseline-grid-ratio}.
\end{proof}

This result propagates an assumed response-error bound; it does not
estimate that bound from the clinical data. It also does not assert
unbiasedness or independence. In particular, changing an event time can
add or remove observed rows from a median window. Such a change violates
the fixed-index-set condition, even if the time shift is small. For the
recorded-point sample-median feature, window-shift analyses must recompute
the included rows and report their counts.

\begin{proposition}[Supported reconstruction with alignment uncertainty]
\label{th:alignment-grid}
Let $t_1<\cdots<t_q$ be the acquisition nodes used to reconstruct
$[6,16]$, including any exterior observations needed to interpolate
its boundaries. Suppose that the ideal signed response $f$ is
$L$-Lipschitz on an interval containing these nodes, their translates
by $h$, and $[6,16]$. At these nodes, let
\begin{equation}
 y_j=b+f(t_j+h)+e_j,
 \qquad |e_j|\leq\epsilon,
 \qquad |\widehat b-b|\leq\eta.
 \label{eq:alignment-observation}
\end{equation}
Assume the observations bracket $[6,16]$ and every acquisition gap
intersecting this interval has length at most $H$. Form $p$ by linear
interpolation of $y_j-\widehat b$ at any missing boundaries $6,12,16$,
followed by clipping all augmented-grid node values and joining them
linearly, exactly as in \eqref{eq:posareas}. Set
\begin{equation}
 \delta_* = \epsilon+\eta+L|h|+\frac{LH}{2}.
 \label{eq:alignment-total-error}
\end{equation}
Then
\begin{equation}
 \sup_{6\leq t\leq16}|p(t)-[f(t)]_+|\leq\delta_*.
 \label{eq:alignment-reconstruction}
\end{equation}
If $D(f)\geq d_0>10\delta_*$, the implemented area fraction satisfies
\begin{equation}
 \left|\frac{\int_6^{12}p(t)\,dt}{\int_6^{16}p(t)\,dt}-E(f)\right|
 \leq
 \frac{\{6(1-E(f))+4E(f)\}\delta_*}{d_0-10\delta_*}
 \leq\frac{6\delta_*}{d_0-10\delta_*}.
 \label{eq:alignment-fraction-error}
\end{equation}
\end{proposition}

\begin{proof}
Put $v_j=y_j-\widehat b$. Then
$|v_j-f(t_j)|\leq\epsilon+\eta+L|h|$.
Fix $t$ in an original acquisition gap $[a,c]$. Let $u\leq t\leq v$
be its neighboring nodes in the augmented grid. For any scalar $z$,
the positive-part Lipschitz inequality gives
$|[x]_+-[z]_+|\leq|x-z|$. Apply this inequality at $u,v$ with $z=f(t)$,
and then use the convex weights from both interpolation stages.
Because composition of linear interpolation within $[a,c]$ preserves
the barycentric weights, this yields
\begin{align}
 |p(t)-[f(t)]_+|
 &\leq \frac{c-t}{c-a}|v_a-f(t)|
        +\frac{t-a}{c-a}|v_c-f(t)|\notag\\
 &\leq \epsilon+\eta+L|h|
       +\frac{2L(t-a)(c-t)}{c-a}
 \leq\epsilon+\eta+L|h|+\frac{LH}{2}.
 \label{eq:alignment-interpolation-proof}
\end{align}
Here $v_a,v_c$ denote the corrected observations at the two original
endpoints. The same bound holds at acquisition nodes by continuity.
Integrating over the nested early and remaining intervals and applying
the ratio identity from Lemma~\ref{th:reconstruction} proves
\eqref{eq:alignment-fraction-error}.
\end{proof}

The terms in \eqref{eq:alignment-total-error} separate observation error,
baseline error, a common alignment shift, and the resolution of the
retained grid. No exponential kinetic model or uniform sampling is
required. The bound is conditional on the stated regularity and error
bounds; it supplies neither a numerical quality threshold nor an
estimated confidence interval. Large or discontinuous switching
artifacts can invalidate a small Lipschitz constant across the relevant
boundary. A common translation also does not describe uncertainty in
the injection duration. These situations require their own declared
sensitivity analyses. The proposition concerns supported reconstruction
of the area fraction and does not replace the discrete retention median
by an interpolated or time-weighted median.

\begin{figure}[!ht]
\centering
\includegraphics[width=0.94\textwidth]{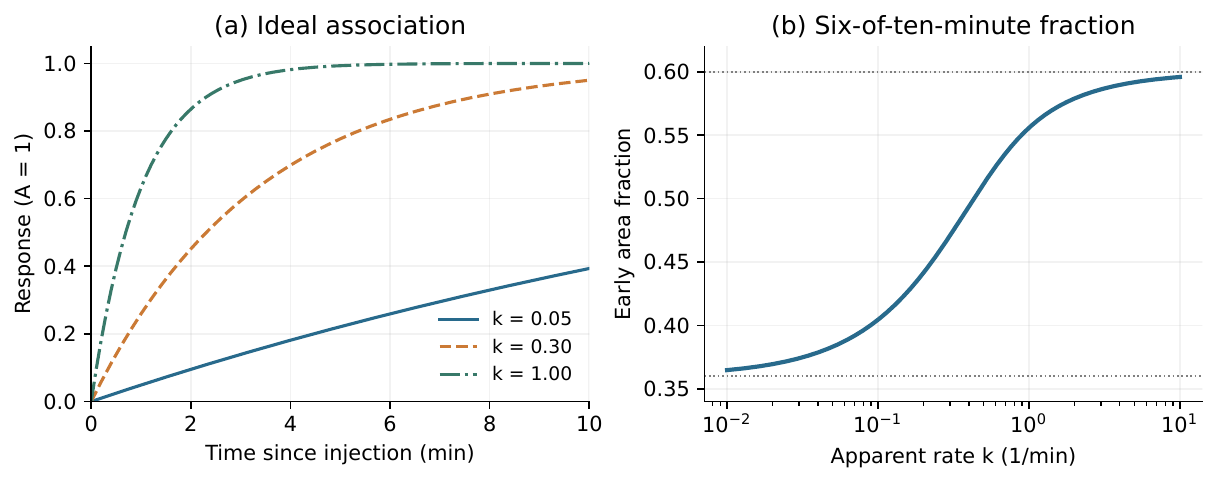}
\caption{Analytical illustration under the ideal exponential association model. Panel (a) varies the apparent rate at unit amplitude; panel (b) shows the six-of-ten-minute area fraction. These are model curves, not patient observations. Both panels are generated by the accompanying Python code.}
\label{fig:idealshape}
\end{figure}
\section{Proofs of the analytical results}
\label{sec:analytical-proofs}

\subsection{Proof of Proposition~\ref{th:slopes}}
\label{p:slopes}
\begin{proof}
Differentiation of \eqref{eq:th-association} gives
$f'(t)=Ak e^{-k(t-6)}$ and hence $f'(6)=Ak$.
Since $f(6)=0$ and $f(12)=A(1-e^{-6k})$, the secant formula follows. Its derivatives with respect to $A$ and $k$ are $(1-e^{-6k})/6>0$ and $Ae^{-6k}>0$, respectively. The corresponding derivatives of $Ak$ are $k>0$ and $A>0$.

For the fixed offsets, put $f_j=A(1-e^{-ku_j})$ and
$S_{uu}=\sum_j(u_j-\bar u)^2>0$. The OLS slope with an intercept is
\begin{equation}
 \frac{\sum_j(u_j-\bar u)f_j}{S_{uu}}
 =-A\frac{\sum_j(u_j-\bar u)e^{-ku_j}}{S_{uu}},
 \label{eq:p-ols-formula}
\end{equation}
because $\sum_j(u_j-\bar u)=0$. To verify positivity, use
\begin{equation}
 \sum_j(u_j-\bar u)f_j
 =\frac1q\sum_{j<\ell}(u_j-u_\ell)(f_j-f_\ell)>0.
 \label{eq:p-ols-positivity}
\end{equation}
Each product is positive for distinct offsets because $u\mapsto A(1-e^{-ku})$ is strictly increasing. Formula \eqref{eq:p-ols-formula} also gives linear scaling in $A$.

For nonmonotonicity in $k$, a valid two-point design has offsets $u_1=1$ and $u_2=6$. Its OLS slope and derivative are
\begin{equation}
 S_{\mathrm{OLS}}(A,k)=\frac A5(e^{-k}-e^{-6k}),\qquad
 \frac{\partial S_{\mathrm{OLS}}(A,k)}{\partial k}
 =\frac A5(-e^{-k}+6e^{-6k}).
 \label{eq:p-ols-counterexample}
\end{equation}
The derivative is positive for $0<k<(\log6)/5$ and negative for $k>(\log6)/5$. This supplies a fixed-design counterexample to general monotonicity.
\end{proof}

\subsection{Proof of Proposition~\ref{th:early-fraction}}
\label{p:early-fraction}
\begin{proof}
With $u=t-6$, direct integration yields
\begin{equation}
 \int_6^{6+a} f(t)\,dt
 =A\int_0^a(1-e^{-ku})\,du
 =A\left\{a-\frac{1-e^{-ak}}{k}\right\}.
 \label{eq:p-association-area}
\end{equation}
The integrand is nonnegative and is positive except at $u=0$. Thus the positive-part and signed areas agree, the denominator is positive, and cancelling $A$ gives \eqref{eq:th-early-formula}.

We prove strict monotonicity for any $0<a<b$ and then take $a=6$ and $b=10$. Define
$g_k(u)=1-e^{-ku}$ and $Q_a(k)=\int_0^a g_k(u)\,du$.
Differentiating the ratio under the integral sign, which is valid on these finite intervals, gives
\begin{equation}
 \frac{d}{dk}\frac{Q_a(k)}{Q_b(k)}
 =\frac{1}{Q_b(k)^2}
   \int_0^a\int_a^b
   \{\partial_k g_k(u)g_k(v)-g_k(u)\partial_k g_k(v)\}
   \,dv\,du.
 \label{eq:p-fraction-derivative}
\end{equation}
Here the contribution with both arguments in $[0,a]$ cancels by antisymmetry. For $u>0$, put
\begin{equation}
 h_k(u)=\frac{\partial_k g_k(u)}{g_k(u)}
       =\frac{u}{e^{ku}-1}.
 \label{eq:p-log-derivative}
\end{equation}
Its derivative with respect to $u$ has the sign of
$e^{ku}-1-ku e^{ku}$.
For $x>0$, the function $e^x(1-x)-1$ is strictly negative, since it vanishes at zero and has derivative $-xe^x<0$. Hence $h_k$ is strictly decreasing on $(0,b]$. The integrand in \eqref{eq:p-fraction-derivative} equals
$g_k(u)g_k(v)\{h_k(u)-h_k(v)\}$ and is strictly positive for $0<u<a<v<b$. This set has positive area, proving strict monotonicity.

As $k\downarrow0$, Taylor expansion gives
\begin{equation}
 Q_a(k)=\frac{a^2k}{2}-\frac{a^3k^2}{6}+O(k^3),
 \qquad
 \frac{Q_a(k)}{Q_b(k)}\longrightarrow\frac{a^2}{b^2}.
 \label{eq:p-small-rate-limit}
\end{equation}
As $k\to\infty$, $Q_a(k)\to a$ and $Q_b(k)\to b$, so the ratio tends to $a/b$. Strict monotonicity and these two limits imply the strict bounds in \eqref{eq:th-early-range}. Finally, $[cf]_+=c[f]_+$ for $c>0$, so both $N$ and $D$ scale by $c$ and $E(cf)=E(f)$ whenever $D(f)>0$.
\end{proof}

\subsection{Proof of Proposition~\ref{th:retention}}
\label{p:retention}
\begin{proof}
Changing variables to $u=t-16$ gives
\begin{equation}
 \mathcal R_{\mathrm{mean}}
 =\frac{R_{16}}4\int_0^4e^{-k_{\mathrm{off}}u}\,du
 =R_{16}\frac{1-e^{-4k_{\mathrm{off}}}}{4k_{\mathrm{off}}}.
 \label{eq:p-wash-mean}
\end{equation}
The derivative with respect to $R_{16}$ is positive. Differentiating under the integral sign also gives
\begin{equation}
 \frac{\partial\mathcal R_{\mathrm{mean}}}{\partial k_{\mathrm{off}}}
 =-\frac{R_{16}}4\int_0^4u e^{-k_{\mathrm{off}}u}\,du<0.
 \label{eq:p-wash-mean-derivative}
\end{equation}

Let $U$ have the uniform distribution on $[16,20]$. Since $f$ is continuous and strictly decreasing,
$\Pr\{f(U)\leq f(18)\}=\Pr(U\geq18)=1/2$ and
$\Pr\{f(U)\geq f(18)\}=1/2$.
The distribution of $f(U)$ has a unique median, namely
$f(18)=R_{16}e^{-2k_{\mathrm{off}}}$.
Its derivatives with respect to $R_{16}$ and $k_{\mathrm{off}}$ are $e^{-2k_{\mathrm{off}}}>0$ and $-2R_{16}e^{-2k_{\mathrm{off}}}<0$.

For ordered sampling times, the responses are ordered in reverse. The central response when $q=2h+1$ is $f(t_{h+1})$; the two central responses when $q=2h$ are $f(t_h)$ and $f(t_{h+1})$. The standard sample-median convention gives \eqref{eq:th-sample-retention}. Each term is a positive multiple of $R_{16}$ and has a nonpositive derivative in $k_{\mathrm{off}}$. This derivative is strictly negative for a term whose time is later than $16$, proving the remaining statements.
\end{proof}

\subsection{Proof of Proposition~\ref{th:replicate-variance}}
\label{p:replicate-variance}
\begin{proof}
For two values, the usual sample median is their arithmetic mean, irrespective of their order. Conditional on $\theta$, the mean of $(X_B+X_F)/2$ is $\theta$. The variance identity for a sum gives
\begin{equation}
 \Var\left(\frac{X_B+X_F}{2}\,\middle|\,\theta\right)
 =\frac14\{\Var(\eta_B\mid\theta)+\Var(\eta_F\mid\theta)
           +2\operatorname{Cov}(\eta_B,\eta_F\mid\theta)\}
 =\frac{\sigma_B^2+\sigma_F^2+2\rho\sigma_B\sigma_F}{4}.
 \label{eq:p-repeat-variance}
\end{equation}
Substituting $\sigma_B=\sigma_F=\sigma$ yields $\sigma^2(1+\rho)/2$. Taking $\rho=0$ gives $\sigma^2/2$; conditional independence is sufficient for this zero correlation under the stated finite-variance assumptions.
\end{proof}

\subsection{Proof of Proposition~\ref{th:fusion-support}}
\label{p:fusion-support}
\begin{proof}
Under the first set of conditions, controls satisfy
$\max(x_1,x_2)\leq a<c$, whereas cases satisfy
$\max(x_1,x_2)\geq b>c$.
Under the second set, controls satisfy $x_1+x_2\leq a<c$ and cases satisfy $x_1+x_2\geq b>c$.
These inequalities establish the respective separation claims.

The caution about a common numerical threshold follows from
\begin{equation}
 \begin{aligned}
 \Pr\{\max(x_1,x_2)>c\mid G=0\}
 &=\Pr\{x_1>c\ \text{or}\ x_2>c\mid G=0\}\\
 &\geq\Pr(x_j>c\mid G=0),\qquad j=1,2.
 \end{aligned}
 \label{eq:p-max-control-probability}
\end{equation}
Thus the support examples provide no unconditional improvement at a specified false-positive rate.
\end{proof}

\subsection{Proof of Lemma~\ref{th:reconstruction}}
\label{p:reconstruction}
\begin{proof}
For an interval $W$,
$|\int_W(g-f)\,dt|\leq\int_W|g-f|\,dt\leq|W|\delta$.
The map $x\mapsto[x]_+$ is $1$-Lipschitz, giving
$|\int_W[g]_+\,dt-\int_W[f]_+\,dt|\leq|W|\delta$ as well.
Dividing by $|W|>0$ proves the average bounds.

In particular, $|\Delta N|\leq6\delta$ and $|\Delta D|\leq10\delta$, so
$D(g)=D+\Delta D\geq d_0-10\delta>0$.
Since $N(f)=ED$, subtracting the two ratios gives the exact identity
\begin{equation}
 \frac{N(f)+\Delta N}{D+\Delta D}-\frac{N(f)}D
 =\frac{\Delta N-E\Delta D}{D+\Delta D}.
 \label{eq:p-ratio-identity}
\end{equation}
To use the nested integration windows, let
$T(f)=\int_{12}^{16}[f(t)]_+\,dt$ and
$\Delta T=T(g)-T(f)$.
Then $D(f)=N(f)+T(f)$, $\Delta D=\Delta N+\Delta T$, and $|\Delta T|\leq4\delta$. Also $0\leq E\leq1$. Consequently,
\begin{equation}
 |\Delta N-E\Delta D|
 =|(1-E)\Delta N-E\Delta T|
 \leq\{6(1-E)+4E\}\delta
 \leq6\delta.
 \label{eq:p-nested-area-bound}
\end{equation}
Combining \eqref{eq:p-ratio-identity} and \eqref{eq:p-nested-area-bound} with the denominator lower bound proves \eqref{eq:th-ratio-bound}.

Subtracting $(\Delta N-E\Delta D)/D$ from \eqref{eq:p-ratio-identity} gives
\begin{equation}
 r=-\frac{(\Delta N-E\Delta D)\Delta D}{D(D+\Delta D)}.
 \label{eq:p-exact-remainder}
\end{equation}
Use $|\Delta N-E\Delta D|\leq6\delta$, $|\Delta D|\leq10\delta$, $D\geq d_0$, and $D+\Delta D\geq d_0-10\delta$ to obtain the stated remainder bound. Finally, if $g_n\to f$ uniformly and $D(f)>0$, fix any $0<d_0\leq D(f)$. The denominator condition holds for all sufficiently large $n$, and all the displayed error bounds tend to zero.
\end{proof}

\subsection{Proof of Lemma~\ref{th:fixed-design}}
\label{p:fixed-design}
\begin{proof}
The OLS slope with an intercept is
$S_{\mathrm{OLS}}(y)=S_{tt}^{-1}\sum_j(t_j-\bar t)y_j$.
Therefore
\begin{equation}
 |S_{\mathrm{OLS}}(y)-S_{\mathrm{OLS}}(z)|
 \leq\frac{\sum_j|t_j-\bar t|\,|y_j-z_j|}{S_{tt}}
 \leq\delta\frac{\sum_j|t_j-\bar t|}{S_{tt}}.
 \label{eq:p-ols-perturbation}
\end{equation}
By the Cauchy--Schwarz inequality,
$\sum_j|t_j-\bar t|\leq\sqrt{qS_{tt}}$, proving the second slope bound.
Adding a constant $c$ to the responses changes the numerator by
$c\sum_j(t_j-\bar t)=0$, establishing baseline-shift invariance.

For the medians, the coordinatewise inequalities
$y_j-\delta\leq z_j\leq y_j+\delta$ imply the corresponding order-statistic inequalities
\begin{equation}
 y_{(\ell)}-\delta\leq z_{(\ell)}\leq y_{(\ell)}+\delta,
 \qquad \ell=1,\ldots,q.
 \label{eq:p-order-statistic-bound}
\end{equation}
For example, at least $\ell$ coordinates of $y$ are at most $y_{(\ell)}$, so the corresponding coordinates of $z$ are at most $y_{(\ell)}+\delta$, proving the upper bound; exchanging $y$ and $z$ proves the lower bound. Taking the central order statistic for odd $q$, or the average of the two central order statistics for even $q$, gives the sample-median bound.
\end{proof}

\section{Source selection and candidate-alignment implementation}
\label{sec:alignment-implementation}
\subsection{Selection can be reconstructed independently of event time}
The source audit covers 91 original eight-column files and 122 selected two-column files retained in the research archive. The source comparison identifies 114 selected files whose observations are fully explained by an original source or its row subset, allowing recorded export rounding. Eight files contain partial value processing, filling, extension, or an additional source segment. These transformations are enumerated rather than silently reversed. Source agreement does not establish an unrecorded biological sample identity.

The group-2 selection retains 34 different source column pairs per marker. Its finite-response sources not retained are PD-L1 C6, C9, C10, H2, and H4 in pair 1; PD-L1 H3 in pair 3; Siglec-15 H4 in pair 1; and Siglec-15 C4, C5, H2, and H3 in pair 3. Siglec-15 H5 pair 3 has entirely missing responses and was also not retained. The source index refers to the original column pairing, not an inferred filename suffix.

\begin{figure}[!ht]
\centering
\includegraphics[width=\textwidth]{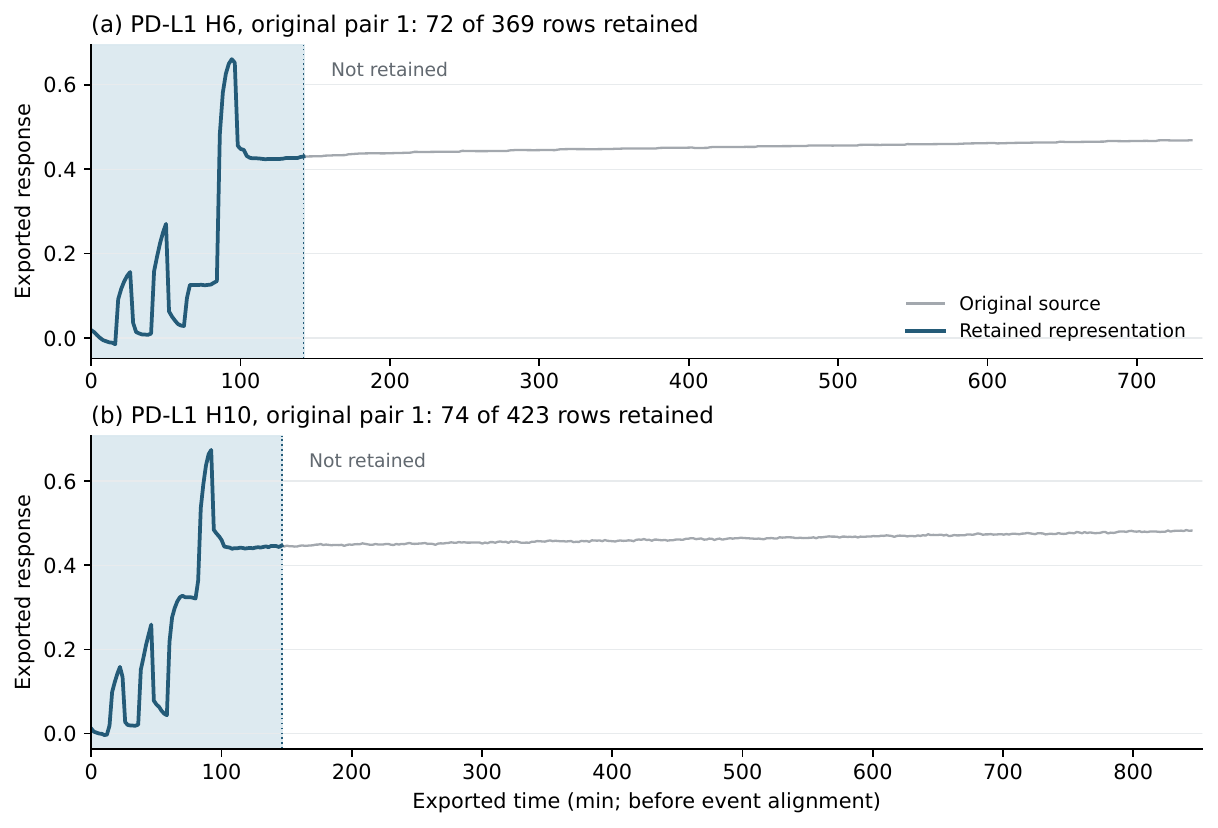}
\caption{Examples of actual within-source selection in group-2 PD-L1 H6 and H10. The original acquisition trace and its retained segment are shown on the original clock. Unretained portions are displayed only to document the prior selection and are not returned to the analysis. The plot does not infer the reason for each excluded row and does not require target-phase alignment.}
\label{fig:retainedsegments}
\end{figure}

\subsection{Candidate events and numerical support}
The event input table retains 69 file entries, including the repeated H2 Siglec-15 export. Deduplication selects one of its entries as canonical before feature aggregation. The PD-L1 C5-F and Siglec-15 C1F entries have no candidate onset/stop bracket. They are not assigned another source's timing, and their patients remain represented through other retained sources. Thus 66 of the 68 distinct sources have candidate events, while the patient-level retained-response comparison includes all 20 study patients.

Response transitions and valve switching are different physical events. In particular, source-specific transport and observed sampling can make paired traces transition at different recorded times. Applying a universal subtraction based on the approximate 90--100\,min protocol, or forcing one source's transition time onto another, would introduce an additional unsupported assumption. The candidate table instead preserves each recorded onset/stop interval and its annotation basis.

When the feasible ten-minute interval in \eqref{eq:eventintersection} is empty, the separate transition midpoints support an explicitly qualified retained-response calculation but not the original fixed-duration early fraction. The implementation checks this duration condition before applying the area-reconstruction guard. Siglec-15 C5B is therefore recorded as ten-minute incompatible: its separate candidate midpoints give a duration of approximately 10.531\,min. Its deleted sequence also leaves an approximately 3.508\,min sampling gap, but that gap is not the recorded cause of unavailability in this execution. A compatible candidate injection interval spanning that gap would additionally fail the 2.5\,min reconstruction guard. These are feature-availability rules applied to the retained data; they do not reverse its quality selection.

The files record observed baseline and retention counts, candidate duration, area support, positive area, and feature availability. Moving a candidate window can change these sets even for a one-minute shift. The fixed-design bounds in Appendix~\ref{sec:analytical-properties} therefore do not justify assuming unchanged median behavior under arbitrary window translations.

\section{Patient-level feature and score table}
\label{sec:patient-table}
\begin{table}[!htbp]
\centering\footnotesize
\caption{Patient-level features and descriptive scores under unshifted candidate alignment. Raw retained-response features use the exported response scale; early fractions are dimensionless. Dashes indicate unavailable features or exclusion from the common-shape set. $R_\Sigma(20)$ uses all-patient scaling, whereas $R_\Sigma(13)$ and $S_{.85}(13)$ use the same 13-patient scaling population. Only the last two columns form the matched shape comparison.}
\label{tab:patientfeatures}
\setlength{\tabcolsep}{3.5pt}
\renewcommand{\arraystretch}{1.0}
\begin{tabular}{lrrrrrrr}
\toprule
ID & $R_{\rm PD}$ & $R_{\rm Sig}$ & $E_{\rm PD}$ & $E_{\rm Sig}$ & $R_\Sigma$ (20) & $R_\Sigma$ (13) & $S_{.85}$ (13) \\
\midrule
C1 & 0.1053 & 0.0909 & 0.531 & -- & -1.010 & -- & -- \\
C2 & 0.0917 & 0.1037 & 0.534 & 0.541 & -1.026 & -1.289 & -0.982 \\
C3 & 0.0068 & 0.1733 & 0.535 & 0.542 & -1.186 & -1.613 & -1.251 \\
C4 & 0.2894 & 0.2103 & 0.551 & 0.547 & 0.864 & 1.144 & 1.258 \\
C5 & 0.3555 & 0.3045 & 0.494 & -- & 1.826 & -- & -- \\
C6 & 0.1069 & 0.0943 & 0.516 & -- & -0.980 & -- & -- \\
C7 & 0.2694 & 0.1007 & 0.525 & 0.523 & 0.114 & 0.280 & 0.107 \\
C8 & 0.1414 & 0.4606 & 0.522 & 0.522 & 1.312 & 1.384 & 1.008 \\
C9 & 0.1653 & 0.4561 & 0.538 & 0.531 & 1.442 & 1.570 & 1.368 \\
C10 & 0.1653 & 0.1899 & 0.539 & 0.536 & -0.060 & -0.093 & 0.014 \\
H1 & 0.0960 & 0.1367 & 0.535 & 0.528 & -0.812 & -1.044 & -0.901 \\
H2 & 0.3047 & 0.2412 & -- & -- & 1.138 & -- & -- \\
H3 & 0.0296 & 0.1706 & -- & -- & -1.053 & -- & -- \\
H4 & 0.0875 & 0.0393 & 0.557 & -- & -1.417 & -- & -- \\
H5 & 0.2756 & 0.2624 & 0.566 & -- & 1.068 & -- & -- \\
H6 & 0.2934 & 0.0607 & 0.570 & 0.510 & 0.045 & 0.245 & 0.284 \\
H7 & 0.1486 & 0.2295 & 0.530 & 0.515 & 0.054 & 0.005 & -0.169 \\
H8 & 0.1615 & 0.2710 & 0.527 & 0.530 & 0.373 & 0.379 & 0.265 \\
H9 & 0.0601 & 0.2275 & 0.565 & 0.514 & -0.533 & -0.798 & -0.610 \\
H10 & 0.1854 & 0.1490 & 0.520 & 0.516 & -0.160 & -0.169 & -0.392 \\
\bottomrule
\end{tabular}

\end{table}


\clearpage
\begin{thebibliography}{99}
\small
\bibitem{wang2025}
Wang Y, Liu Y, Lee J, Chen H. Ligand engineering-mediated plasma/supramolecular self-assembly of superstructures sensitized dual immune checkpoint biosensors. \emph{Chemical Engineering Journal}. 2025;522:167712. \url{https://doi.org/10.1016/j.cej.2025.167712}.

\bibitem{huang2021pdl1}
Huang X, Zhang ZH, Chen J, Mao Z, Zhu H, Liu Y, Zhu Z, Chen H. One dimensional magneto-optical nanocomplex from silver nanoclusters and magnetite nanorods containing ordered mesocages for sensitive detection of PD-L1. \emph{Biosensors and Bioelectronics}. 2021;189:113385. \url{https://doi.org/10.1016/j.bios.2021.113385}.

\bibitem{huang2021sig}
Huang X, et al. Magnetic field-aligned Fe$_3$O$_4$-coated silver magnetoplasmonic nanochain with enhanced sensitivity for detection of Siglec-15. \emph{Biosensors and Bioelectronics}. 2021;191:113448. \url{https://doi.org/10.1016/j.bios.2021.113448}.

\bibitem{grasmeier2023}
Grasmeier MK, Weber S, Treiber M, Thaler MA, Luppa PB. Surface plasmon resonance assays for the therapeutic drug monitoring of infliximab indicate clinical relevance of anti-infliximab antibody binding properties. \emph{Clinical Chemistry and Laboratory Medicine}. 2023;61(7):1255--1265. \url{https://doi.org/10.1515/cclm-2022-0949}.

\bibitem{vollenberg2019}
Vollenberg R, et al. Glycoprotein V is a relevant immune target in patients with immune thrombocytopenia. \emph{Haematologica}. 2019;104(6):1237--1243. \url{https://doi.org/10.3324/haematol.2018.211086}.

\bibitem{metayer2025}
Metayer I, Forest-Nault C, Guimond J, Joubert S, Henry O, Durocher Y, De Crescenzo G, Gaudreault J. A surface plasmon resonance-based integrated assay for quantification and glycosylation characterization of monoclonal antibodies in crude heterogeneous samples. \emph{Biotechnology and Bioengineering}. 2025;122(10):2724--2738. \url{https://doi.org/10.1002/bit.70016}.

\bibitem{volpe2025}
Volpe J, et al. Interpretating SPR-derived reaction kinetics via self-organizing maps for diagnostic applications. \emph{ACS Sensors}. 2025;10(10):8129--8139. \url{https://doi.org/10.1021/acssensors.5c03250}.

\bibitem{carroll2016}
Carroll J, Raum M, Forsten-Williams K, T\"auber UC. Ligand-receptor binding kinetics in surface plasmon resonance cells: A Monte Carlo analysis. \emph{Physical Biology}. 2016;13(6):066010. \url{https://doi.org/10.1088/1478-3975/13/6/066010}.

\bibitem{cohen2016}
Cohen JF, et al. STARD 2015 guidelines for reporting diagnostic accuracy studies: explanation and elaboration. \emph{BMJ Open}. 2016;6(11):e012799. \url{https://doi.org/10.1136/bmjopen-2016-012799}.

\bibitem{pepe2008}
Pepe MS, Feng Z, Janes H, Bossuyt PM, Potter JD. Pivotal evaluation of the accuracy of a biomarker used for classification or prediction: standards for study design. \emph{Journal of the National Cancer Institute}. 2008;100(20):1432--1438. \url{https://doi.org/10.1093/jnci/djn326}.
\bibitem{nistbinomial}
NIST/SEMATECH. Confidence intervals for a proportion. In: \emph{e-Handbook of Statistical Methods}, Section 7.2.4.1. \url{https://www.itl.nist.gov/div898/handbook/prc/section2/prc241.htm}. Accessed 10 September 2026.
\end{thebibliography}
\end{document}